\documentclass[12pt,preprint]{aastex}

\def \minpoint {${.}\!\!^{\prime}$}
\def\arcsec{$^{\prime\prime}$}
\def\arcmin{$^{\prime}$}
\def\ltsim{ \,{}^<_\sim\, }

\shorttitle{GCs and X-ray Point Sources in NGC 5128}
\shortauthors{K.~A.~Woodley et al.}
\begin{document}

\title{Globular Clusters and X-ray Point Sources in Centaurus~A (NGC 5128)}

\author{Kristin A.~Woodley\altaffilmark{1}, 
Somak Raychaudhury\altaffilmark{2,3},
Ralph P.~Kraft\altaffilmark{2},
William E.~Harris\altaffilmark{1}, 
Andr\'{e}s Jord\'{a}n\altaffilmark{2,4},
Katherine E. Whitaker\altaffilmark{2,5},
Christine Jones\altaffilmark{2},
William R. Forman\altaffilmark{2}, 
Stephen S. Murray\altaffilmark{2}
}
\altaffiltext{1}{Department of Physics \& Astronomy, McMaster
University,
  Hamilton ON  L8S 4M1, Canada; woodleka@physics.mcmaster.ca, harris@physics.mcmaster.ca}
\altaffiltext{2}{Harvard-Smithsonian Center for Astrophysics, 60 Garden
Street,
  MS-67, Cambridge, MA 02138, USA; kraft@head.cfa.harvard.edu,
  ajordan@cfa.harvard.edu,  
cjf@head.cfa.harvard.edu, wrf@head.cfa.harvard.edu, 
ssm@head.cfa.harvard.edu}
\altaffiltext{3}{School of Physics \& Astronomy, University of
Birmingham, Birmingham B15 2TT, UK; somak@star.sr.bham.ac.uk}
\altaffiltext{4}{European Southern Observatory, 
Karl-Schwarzschild-Str.\ 2 85748
Garching bei M\"{u}nchen, Germany}
\altaffiltext{5}{Department of Astronomy, Yale University,
260 Whitney Avenue, New Haven, CT 06511, USA; katherine.whitaker@yale.edu}


\begin{abstract}
We detect 353 X-ray point sources, mostly low-mass X-ray binaries
(LMXBs), in four {\em Chandra} observations
of Centaurus~A (NGC 5128), the nearest giant early-type galaxy,
and correlate this point source population
with the largest available ensemble of confirmed and likely 
globular clusters associated with this galaxy.
Of the X-ray sources, 31 are coincident with 30 globular clusters 
that are confirmed members of the galaxy by radial velocity measurement (2 X-ray 
sources match one globular cluster within our search radius), while 1 X-ray 
source coincides with a globular cluster resolved by HST images.
Another 36 X-ray point sources match probable, but
spectroscopically unconfirmed, globular cluster candidates.  
The color distribution of globular clusters and cluster candidates
in Cen A is bimodal, and the probability that a red, metal rich
GC candidate contains an LMXB 
is at least 1.7 times that of a blue, metal poor one.
If we consider only spectroscopically confirmed GCs, 
this ratio increases to $\sim$3. 
We find that LMXBs appear preferentially in more luminous (massive) GCs.
These two effects are independent, and the latter is likely a consequence of
enhanced dynamical encounter rates in more massive clusters which have
on average denser cores.  
The X-ray luminosity functions of the LMXBs found in GCs and of
those that are unmatched with GCs reveal similar underlying
populations, though there is some indication that fewer 
X-ray faint LMXBs are found in globular clusters than X-ray bright ones.
Our results agree with previous observations of the connection
of GCs and LMXBs in early-type galaxies and extend previous work 
on Centaurus~A. 
\end{abstract}


\section{Introduction}
\label{sec:intro}

Observations of early-type galaxies with the {\em Chandra} X-ray
Observatory have shown that a significant fraction of low-mass X-ray
binaries (LMXBs) in these galaxies are associated with globular
clusters (GCs) \citep[see][for a review]{fab06}. 
The fraction of LMXBs
identified with known GCs around massive early-type galaxies varies
from at least 20\% in NGC~4697 \citep{sarazin01} to a remarkable
70\% in NGC~1399, the central giant elliptical of the Fornax cluster
\citep{angelini01}.  In the Milky Way $\sim$10\% of all bright LMXBs
are found in GCs, even though the GCs account for $<10^{-3}$ of the stellar mass of the galaxy 
\citep[e.g.][]{katz75,grindlay93}. 

The overabundance of LMXBs in GCs is thought to be due to their
high central densities, which result in greatly enhanced rates
of dynamical interactions with respect to the field.
These interactions can lead to the creation of LMXB progenitors via
dissipative encounters between neutron stars and ordinary stars 
\citep[e.g.][]{fpr75}, or
encounters in which a compact object replaces a member of a binary in
a three-body interaction \citep[e.g.][]{hills76}. This picture
is supported by observations in the Milky Way and nearby early-type
galaxies \citep{pooley03, jordan04, siv07, jordan07} which show that estimates of 
GC dynamical collision rates are good predictors for the presence of 
LMXBs in GCs. 

It has even been suggested \citep{white02} that almost all LMXBs are
formed in the cores of GCs and
either are later ejected from their host clusters 
due to the evaporation of the GCs or escape the cluster due to dynamical 
processes or kick velocities  imparted at birth. Recent analyses 
suggest that this is not the case though, as observations seem to 
require the presence of a bona fide population of field LMXBs besides a 
component whose formation is connected to GCs
\citep{irwin05,juett05}.  Moreover, \cite{fabbiano07} have recently
shown that GCs in NGC 3379 lack low luminosity LMXBs when compared to
the field population, implying that the latter cannot have originated
from GCs \citep[see also ][]{voss07}.
To date, large populations of LMXBs have been
found in X-ray observations of E/S0 galaxies over a wide range of
luminosities \citep[e.g.][]{siv07,kundu07} and in large disk galaxies such as
M31, NGC~3115, and NGC~4594 \citep{fan05,kundu03,distefano03}.

In this paper, we present results from a correlation of the LMXB
population, detected as point sources in Chandra/ACIS observations,
with the optically detected GC population in the galaxy Centaurus~A
(Cen~A, NGC 5128). A previous study of the connection between
GCs and LMXBs in Cen~A, using smaller optical and X-ray catalogs,
is presented in \citet{m04}.
Cen~A is the nearest massive early-type galaxy
($M_B\! =\! -21.1$) \citep{dufour79}, and has been widely studied across the
electromagnetic spectrum \citep[for a comprehensive review,
see][]{israel98}.  X-ray observations of Cen~A
\citep{feig81,turner97,kra00,kra02,kra03a,kra03b,hardcastle07} reveal several
distinct emission components, including a bright, compact and variable
nucleus, an X-ray jet perpendicular to the dust lane and aligned with
radio features, and diffuse emission representing the hot phase of the 
interstellar medium.  In addition,
several hundred bright point sources are seen in X-rays
\citep{kra01,vg06}, mostly representing the LMXB population.

Studying Cen~A has major advantages over studies of the
LMXB/GC connection in other early-type galaxies.  
First, Cen~A is $\sim$5
times closer than the often-studied galaxies in the Virgo and Fornax
clusters, which allows for a much deeper census and better characterization
of its X-ray binary population.  Second, at the distance of Cen~A, a scale of $1^{\prime\prime}$
corresponds to a linear distance of just 18~pc, which is comparable to
the physical diameters of GCs.  Furthermore, Cen~A has a
moderately large GC population \citep[$\simeq 1500$ clusters;
see][]{harris06}, with a bimodal distribution in color, 
equally split between red and blue.

An intriguing aspect of the observations linking GCs
with X-ray point sources has been that LMXBs in
early-type galaxies are more likely to be found in the redder 
(higher-metallicity $Z$) GCs.  This trend already can
be seen in the small Milky Way sample of
LMXBs, but becomes much more evident in larger galaxies
\citep[e.g.][]{angelini01, kundu02,distefano03,jordan04,kim06,siv07,poss07,kundu07}. 
The physical mechanisms underlying this 
observational result are still unknown, 
although several possibilities have been put forward, including metallicity 
dependent variations in the initial mass function \citep{grindlay87}, the link 
between metallicity and the outer convective zones of stars \citep{ivanova06}
and the effects of metallicity on radiation induced stellar winds \citep{maccarone04}.

A recent spectroscopically based survey of the
cluster ages ($\sim 150$ GCs with Lick index measurements) by
\citet{beasley07} indicates that the great majority of the GCs are
old, with ages $\tau \gtrsim 10$ Gyr.  However, perhaps 20\%
of them (almost all on the metal-rich side) scatter to younger
ages in the 6 to 8 Gyr range.  For comparison, \citet{rej05} show that
the field halo stars in Cen~A have a mean inferred age of
$8^{+3}_{-3.5}$ Gyr.  In the Milky Way, the metal-rich
clusters, including the LMXB hosts, are mostly old systems
\citep{rosenberg99}.
The two giant galaxies M87 and M49 
in Virgo are found to have mostly old GC systems 
\citep{cohen98,jordan02,puzia99,beasley00} and also show a
preference for metal-rich GCs to host LMXBs \citep{siv07}.
Thus the preference of X-ray binaries to be in higher metallicity clusters
does not seem to be associated with age and must be 
primarily due to metallicity \citep[see also][]{kundu03}.

Recent work by \cite{jordan07} exploits
the distance to Cen~A in order to explore the detailed
dependence of the incidence of LMXBs in GCs on the structural
parameters of GCs such as their half-light radii and central
densities.  \cite{jordan07} conclude that neither concentration nor
mass are fundamental variables in determining the presence of LMXBs in
GCs, and that the more fundamental parameters relate to central
density and size.

Various studies indicate that the half-light radii of metal-rich
GCs in large E galaxies are $\sim 20\%$ smaller than the
metal-poor clusters (see \cite{kundu01}, \cite{jordan05}, and \cite{gw07} among
others).  \cite{jor04} shows that this could be a result of mass
segregation and the longer lifetimes of lower metallicity stars for a
given mass, under the assumption that the half-mass radii distribution
does not depend on metallicity. In this case, we expect dynamical
effects to be largely decoupled from the metallicity of GCs, but
if the half-light radii reflect indeed a difference in half-{\it mass}
radii at least part of the preference of LMXBs for metal-rich GCs
could be due to their smaller average size and therefore 
denser cores.
It is easy to show that dynamical processes
cannot be responsible for all of the observed effect of metallicity
on the incidence of LMXBs, even if there was a size difference consistent
with that observed in other early-type galaxies. 
Assuming encounter rates scale with 
half-light radii $r_h$ as $r_h^{-2.5}$ \citep[see ][]{siv07} metal-rich
GCs would need to be $\sim 50\%$ smaller than their metal-poor counterparts
in order to fully account for a threefold enhancement of the number
of LMXBs in metal-rich GCs. This level of size difference is ruled out
by the observations.  The metallicity itself seems to be an
important factor in producing LMXBs.

This paper is organized as follows.  We introduce the
observations and discuss data reduction and point source detection in
\S\ref{s:obs}.  In \S\ref{s:matching}, we 
identify GCs from optical observations
which are associated with the X-ray point sources, and discuss their
properties in \S\ref{s:gcmatches}.  In \S\ref{s:xraypts}, we
examine the X-ray properties of the point sources that are associated
with GCs and compare with those not
found in GCs.  Finally, results are summarized in \S\ref{s:remarks}.
In this analysis, we adopt a distance of $3.8 \pm 0.2$ Mpc for 
Cen~A \citep[an average of five precise standard candles including
TRGB, PNLF, SBF, LVPs, and Cepheids; see][and references
therein]{mcl07, ferrarese07} and an 
integrated optical luminosity $M_V^t = -22.0$ \citep{dufour79}.

\section{X-ray Observations}
\label{s:obs}

\subsection{Chandra/ACIS data: Preparation and Analysis}

We first present the analysis of four ACIS observations of Centaurus~A with
the {\em Chandra} X-ray Observatory.
A summary of the observation log is given in
Table~\ref{cxoobslog}.  All observations were made in FAINT mode.  The
level~1 events files of all four data sets were reprocessed to remove
the pixel randomization, apply CTI correction, and remove flaring and
hot pixels.  The work described here includes a
re-processing/re-analysis of our results presented in \citet{kra01}.
For every ACIS observation, X-ray light curves were extracted for each
CCD in the 5.0--10.0 keV band, excluding the nucleus and point
sources, to search for background flares.  All periods when the
background exceeded the mean by 3$\sigma$ were removed.
Total good times of the ACIS-S and ACIS-I observations are $\sim$94 ks
and $\sim$68 ks, respectively.

The four data sets were aligned relative to each other by comparing
the positions of 30 bright point sources within 3$^\prime$ of the
nucleus.  The coordinates of the data sets were adjusted to minimize
the root mean square separations of the ensemble of sources.  The
largest adjustment made to the astrometry for any data set was
1.2\arcsec.  We also investigated whether any of the data sets were
offset in roll angle, but found that no adjustment to the roll was
necessary.  These four data sets were then aligned on the sky in
absolute coordinates by co-aligning the X-ray and radio nuclei.  We
estimate that the four data sets are aligned relative to each other to
better than 0.1\arcsec, and in absolute sky coordinates to better than
0.5\arcsec.

\subsection{Point Source Detection}
\label{s:detection}

The detection of point sources was performed on five data sets,
individually on each of the four observations, and on the combined
events file of the 2002 and 2003 ACIS-S observations (OBSID 2978 \& 3965).
In general, we found it preferable to perform point source detection
individually on each of the four data sets.  Any improvements in
signal to noise by combining all four data sets would be more than
offset by the complexity introduced by the process.  However, since the AO-3
and AO-4 observations were made with the same detector (ACIS-S) and
with nearly identical roll angles, we have used the combination of
these two observations.

For each of the five data sets, we created images at full resolution
(i.e. 1 pixel $\equiv$ 0.492\arcsec) in three energy bands for each CCD,
namely soft (0.5--2.0 keV), hard (2.0--5.0 keV) and combined (0.5--5.0
keV) bands.  In all cases, the background far exceeds the nominal ACIS
particle background over much of the FOV.  The background in the soft
band images is dominated by emission from hot gas in Cen A.  The
morphology of this emission has a complex spatial dependence, 
and the sensitivity is also
spatially modulated by the variable absorption of the dust lane.  The
background in the hard band is dominated by the wings of the point spread
function (PSF) of the bright active nucleus.  This is most significant 
within $\sim$1$'$ of the nucleus, but the X-ray AGN is so bright that it 
adds significantly to the background over the entire ACIS FOV.

Point source detection was performed with the CIAO software package
{\em wavdetect} on spatial scales of 1.0, 1.414, 2.0, 2.828, 4.0,
5.656, 8.0, 11.312, 16.0 pixels, with a threshold of 10$^{-6}$.
Source detection was performed separately on each chip.  We expect to
detect roughly one false source per CCD at the sensitivity limit.
We detect 353 sources, of which $\sim$10 are
statistically likely to be mis-identified fluctuations in the
background.  The spatial distribution of sources is strongly peaked
toward the center of the galaxy, demonstrating that the vast majority
lie within Cen~A and not unrelated foreground or
background objects.  The limiting sensitivity of these observations,
however, is a widely non-uniform function of position on the sky, due to
the spatial distribution of the ISM, differential absorption of the dust
lane, the scatter from the nucleus, and the complex spatial
dependence of the telescope PSF.  Point sources detected in the two
ACIS-I observations are complete and unbiased (defined as a 4$\sigma$
measurement of the luminosity) to a luminosity of 2$\times$10$^{37}$
ergs s$^{-1}$ in the 0.1-10.0 keV band (unabsorbed).  The limiting
sensitivity of the two ACIS-S observations is roughly a factor of two
lower than this because of the longer exposure, greater sensitivity of
the backside illuminated CCD, and the position of the telescope best
focus at the nucleus.

We performed a count rate to luminosity conversion assuming a 5~keV
bremsstrahlung spectrum, typical of Galactic X-ray binaries,
and absorption ($N_H$=8.0$\times$10$^{20}$
cm$^{-2}$) by foreground gas in our galaxy.  The
conversions for each CCD in each AO were computed with the CXC
program PIMMS.  All count rates were exposure corrected and background
subtracted.  The count rate was estimated from the broad band
(0.5--5.0 keV) image, even if the source was detected in only one or
two of the bands.  The source background in each band was determined using the
residual background image from {\em wavdetect}.  
We also fit bremsstrahlung and power law models to
all sources with more than 50 counts in any observation.  With a few
exceptions of heavily absorbed sources, the luminosities derived by
the two methods are in general statistically consistent.

We will describe the spectral properties of the ensemble of sources in
detail in a future publication.  The photometric properties of a
subset of these point sources, based on some or all of these
observations, have been discussed elsewhere by \citet{kra01} and \citet{vg06}.
For the remainder of this paper, we concentrate on describing the
identification of the X-ray point sources with the GCs in the galaxy
and the optical and X-ray properties of the matched GC-LMXB pairs.

\section{X-ray Point Sources in Globular Clusters}
\label{s:matching}

In the three decades since the breakthrough discovery of
just one of its GCs by \cite{gp80}, 415 GCs have been
found in Cen~A, through either radial velocity measurement or high
resolution imaging.  Cen~A, with a low Galactic latitude of
$b=19^\circ$, is in a field contaminated by a large number of
foreground stars. There are also a large number of more distant
galaxies in the field, since our target is in the foreground of the
Hydra-Centaurus and Shapley superclusters \citep{ray89}.  Many
contaminant stars and galaxies have colors similar to those of
GCs \citep[e.g.][]{hhg04II}.  While the use of colors,
morphologies, and sizes of objects all help to reduce the
contamination from the field and select GC candidates,
none are definitive methods for classification.  Radial velocity
measurements and high-resolution imaging (particularly with {\it HST}) are
the best ways to confirm an optically-identified object as a definite
GC in this challenging galaxy.

In this analysis, we use the complete catalog of 415
identified GCs published in
\citet{woodley07}.  Of these, 340 have radial velocity measurements.
The systemic velocity of Cen~A is $541\pm7$ km s$^{-1}$
\citep{hui95}, and its GCs have radial velocity
measurements in the range 200--1000 km s$^{-1}$
\citep[see][]{pff04I,WHH05,woodley07}.  We are currently using
roughly one-quarter to one-third of the entire estimated GC 
population of Cen~A.

The X-ray point sources, found in the Chandra ACIS observations
described in the previous section, were matched to our list of 340
GCs confirmed by radial velocity.  We found 30 of these radially
velocity confirmed GCs matched at least one X-ray source, a
fraction of $8.7\%$, within a search radius of 1.5\arcsec, and 26 of
these fell within a search radius of 1\arcsec.  Of the 75 GCs in
the \cite{woodley07} catalog that are resolved by HST images, only 1
(GC0141) matched an X-ray source.  This brings the matching fraction
down to $7.5\%$ for the total of 415 GCs.  This range of search
radius was chosen to be consistent with the accuracy of the positions
of the GCs, determined from optical images, of
$\pm$ 0.2\arcsec--0.3\arcsec \citep{hhg04II}, as well as our X-ray
point source positioning, which is better than 1\arcsec.  These 30
GC X-ray sources, listed in Table~\ref{tab:GCmatches},
represent a three-fold improvement over the \cite{m04} study
which found 11 matches to {\it confirmed} GCs.

Many additional GC {\sl candidates} (not yet confirmed by velocity
measurement or HST imaging) have been found in previous studies 
based on colors and
morphologies alone \citep[see][]{hhg04II,gomez06}, as well as objects
known to match X-ray sources in previous studies \citep{m04,pff04I}.  
Of these over $\sim 450$ GC candidates, we 
found 38 further matches with X-ray point sources within
a search radius of 1.5\arcsec, all of which also match within 1\arcsec.
These objects also are listed in Table~\ref{tab:GCmatches}.  Because
of the close physical association between LMXBs and GCs established
in other galaxies, we regard these additional matches as very likely to
be genuine GCs.  In total, we find 67 probable
GC X-ray sources in Cen~A, doubling the number presented by
\cite{m04}.

\begin{figure}
\includegraphics[width=0.5\hsize]{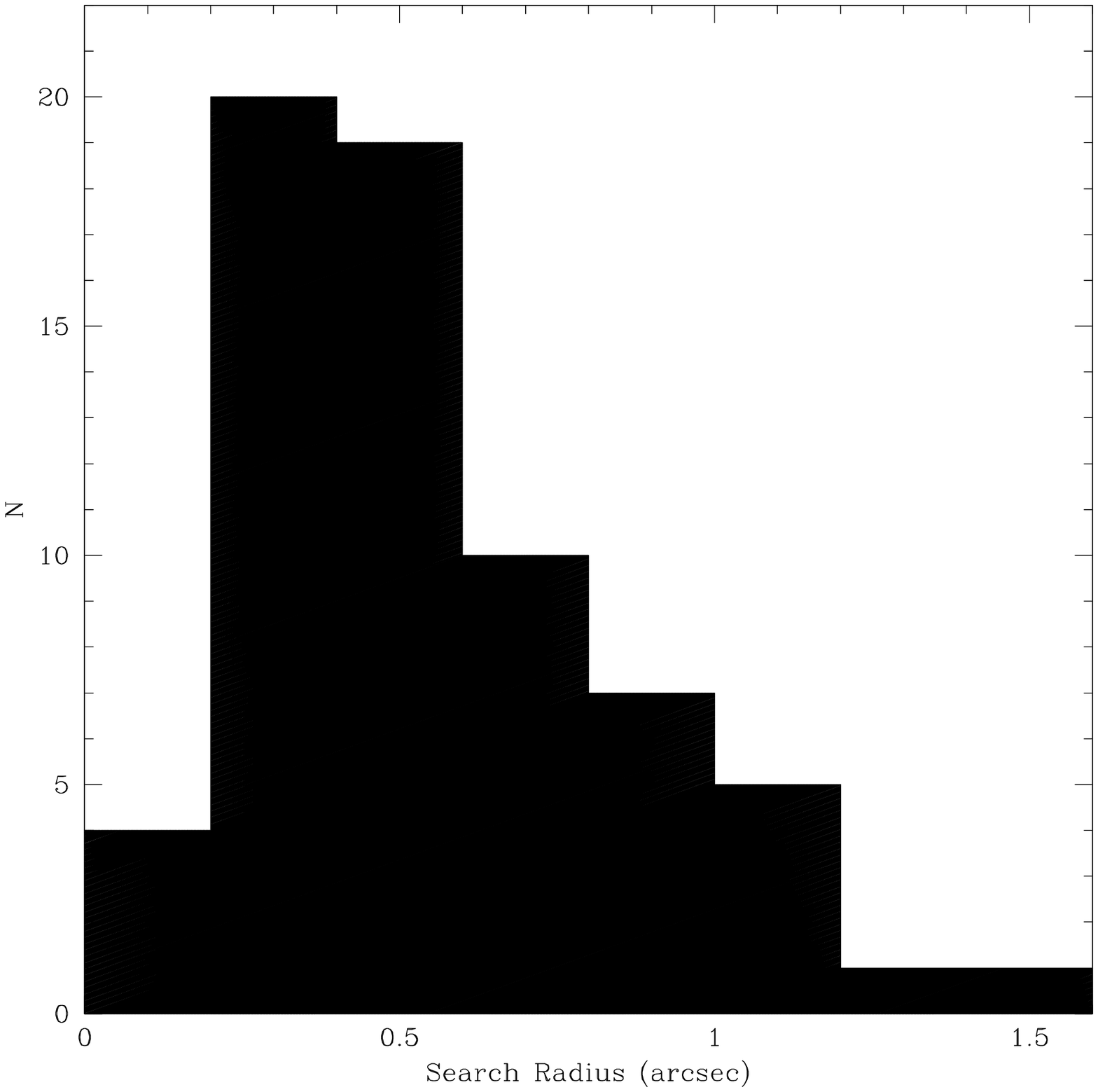}
\includegraphics[width=0.5\hsize]{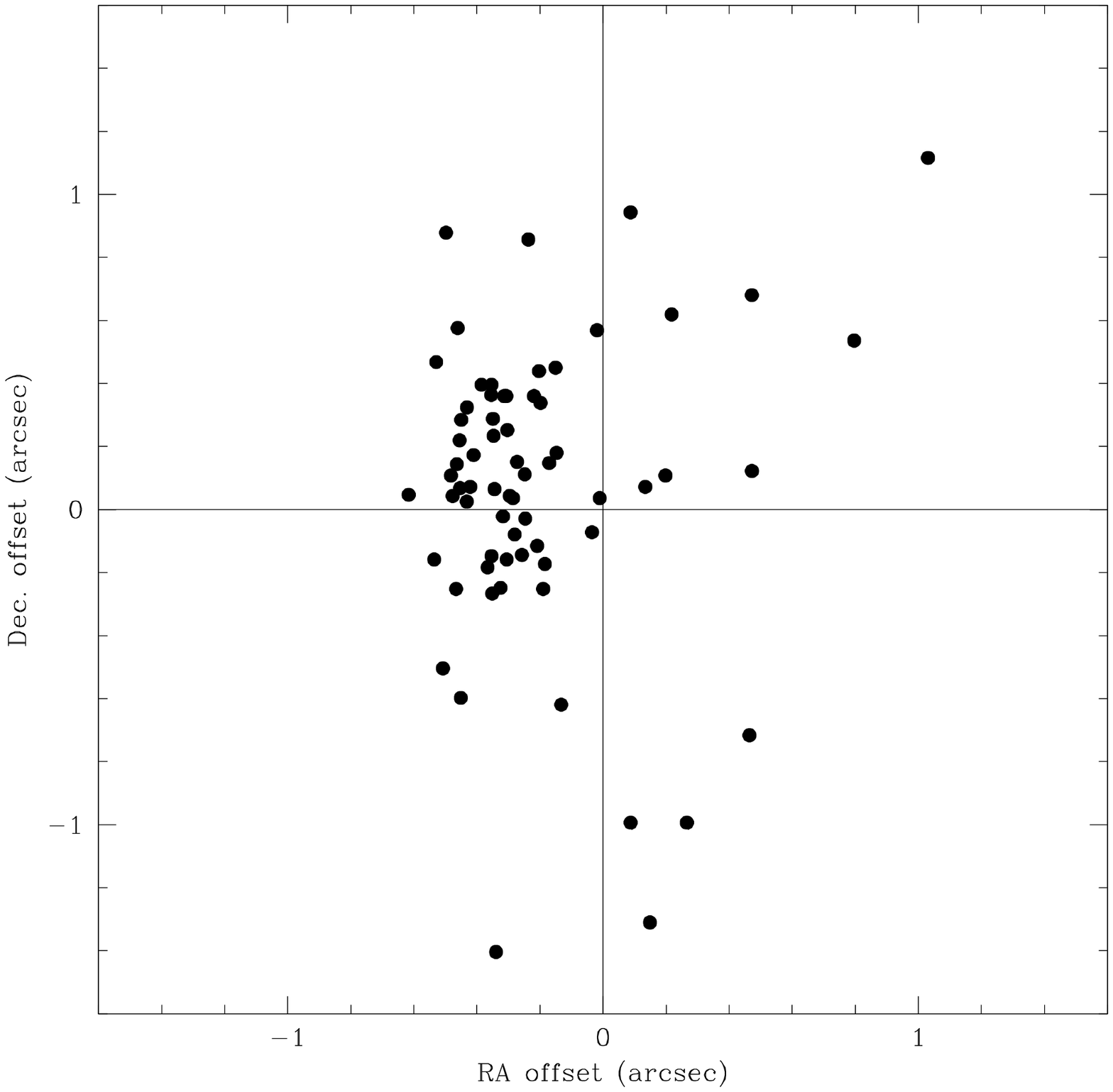}
\caption{(a) The number of X-ray point sources matching GCs binned
  within search radii of 0.2\arcsec.(b) The RA and Dec. offsets 
  of the 67 matched X-ray point sources to GCs.  A
  mean offset between the two catalogs of RA = -0.2026\arcsec and
  Dec. = 0.0690\arcsec exists.  We have used these offsets to align
  the X-ray point source catalog with the GC positions in order to
  search for LMXBs.}

\label{fig:matches}
\end{figure}

Figure~\ref{fig:matches} displays the number of X-ray point sources
that match the GCs within bins of 0.2\arcsec in search radius.  We clearly
see that most matches are within 0.2\arcsec-0.6\arcsec.  This trend is
also seen in the offsets (in arcseconds) in RA and Dec. 
between the 67 matches that we
found in this study, also shown in Fig.~\ref{fig:matches}.  
The small offset between the
two catalogs is well within our outer search radius of 1.5\arcsec.  
However, we shifted the X-ray point sources by the mean offset 
of the matched RA and Dec. positions (RA =-0.2026\arcsec\ and 
Dec. = 0.0690\arcsec) and rematched the sources to all GCs and 
GC candidate objects. We obtained the same result.

We have performed tests on our matching program to determine the
probability of false matches within our sample, assuming a search radius of
1.5\arcsec.  Shifting the positions of the X-ray
sources by 2\arcsec, 3\arcsec, 5\arcsec, and 10\arcsec\ in right
ascension and declination lead to 0, 2, 1, and 1 false matches,
respectively.  We conclude that at most $1-2$ false matches probably exist in
our list of 67 matches.

Although it is unusual to find more than one bright LMXB source within a
single cluster \citep{siv07,kundu07}, it is certainly known to 
occur (e.g. M15 in the Milky Way).
One GC, GC0233, was
found to match two X-ray point sources within 1.5\arcsec, one of them
being within a smaller radius of 0.5\arcsec.
It is possible that both X-ray sources are associated with this GC.

In Table~\ref{tab:GCmatches}, the first column lists the GC 
ID from \citet{woodley07}, where
clusters confirmed by either radial velocity or {\it HST} 
imaging have names beginning with ``GC''.  The following columns are
right ascension and declination (J2000), galactocentric
radius in arcminutes, the various photometric indices ($B$, $V$, $I$,
$T_1$, $(C-T_1)$, when known; magnitudes are not de-reddened), 
and metallicity [Fe/H]. The
$BVI$ data are taken from \cite{pff04I} and the Washington photometry
data is from \cite{hhg04II}.  The metallicity [Fe/H] was determined
from the color $(C-T_1)_0$, using a conversion derived by \cite{hh02},
which produces an uncertainty in [Fe/H] of $\pm0.2$ dex at the
metal-poor end and $\pm0.07$ dex at the metal-rich end for a typical
photometric uncertainty of $\sigma(C\!-\! T_{1}) \simeq 0.1$,
assuming a mean age of $\tau \sim 13$ Gyr.

The positions of the GCs and X-ray sources are shown in
Figure~\ref{fig:positions}, superposed on an optical image, and their
radial positions in Figure~\ref{fig:BV_rad}.  The GC X-ray
sources are well within 10\arcmin\ from the center of Cen~A, 
limited by the extent of the Chandra fields. 

Figure~\ref{fig:x-positions}
is an adaptively smoothed, exposure corrected
image from a combination of the observations
tabulated in Table~1, in the 
the 0.5-2.0 keV band.  The background has not been subtracted.
The positions of the confirmed and candidate GCs are shown
with the same color code as the previous figure.

While the GC
system extends detectably farther out \citep{hhg04II}, the
majority are well within 15\arcmin\ from the galaxy center.
Thus if the X-ray point sources extend beyond the innermost regions
of the galaxy, we would expect only a few more detections ($\ltsim$7)
beyond 15\arcmin, assuming a matching percentage of
$8.7\%$ as we found above. Nevertheless, the outer regions of Cen~A
have not been searched as thoroughly as the inner regions for GCs, 
especially with spectroscopic confirmation.  Similarly, the
inner 5~kpc of the galaxy has very few spectroscopically confirmed
clusters due to the obscuration of the large dust lane extending
across the center.

To estimate the number of GCs obscured in the vicinity
of the dust lane, we show in Figure~\ref{fig:dustlane} the radial
distribution of X-ray detections, all GCs, and GCs 
with associated X-ray sources in the dust lane and its
vicinity. The dust lane is bounded roughly by an ellipse with
semimajor axis of 4\minpoint 2 and axial ratio 0.5 \citep{hhg04II}.
Within this region, there are 119 X-ray point sources and only 17
known GCs, 8 of which have radial velocity confirmation,
many fewer clusters than the estimated $110\pm27$ \citep{hhg04II}.
The fraction of GCs in the 1\arcmin\ annulus
outside the dust lane region identified with LMXBs is $0.28$.
Applying this factor to the estimated number of clusters in the dust
lane predicts $31\pm8$ matches, higher than the 12 matches we found.
In other words, another $\sim 20$ GCs with
LMXBs may remain to be optically identified in this inner region,
though identifying them will be a considerable challenge.

\section{Properties of the Globular Clusters matched with X-ray
Sources}
\label{s:gcmatches} 

\subsection{Optical colors}

Similar surveys of other early-type galaxies show that LMXBs are $\approx$3
times more likely to be found in the redder, more metal-rich GCs. 
These include the studies of M87 \citep{jordan04}, NGC~4636
\citep{poss07}, NGC~1399 \citep{angelini01}, NGC~4472 \citep{kundu02},
and NGC~1553, NGC~4365, NGC~4649, NGC~4697 \citep{sarazin03, siv07} and 
samples of five \citep{kundu07} and eleven \citep{siv07} early-type 
galaxies within $\sim 25$  Mpc.
In Cen~A, \citet{m04} found that LMXBs also are found preferentially
in redder GCs.

The distributions of the optical colors ($(B-V)$, $(V-I)$, and
$(B-I)$) of the Cen~A GCs, and that of their derived 
metallicity [Fe/H], are shown in Figure~\ref{fig:distributions} for
our Cen~A GC sample.  The entire GC population, shown as the open
histogram, shows clear bimodality in all three colors.  The GCs with
associated X-rays, shown as the solid histogram, clearly show a 
preference for the redder GCs, as they do in other
similar early-type galaxies.  
The hatched histogram includes all GCs
within the same radial range as the outermost X-ray point source
(corresponding to 16.5 kpc), and includes 314 out of 415 GCs.

The raw color-magnitude diagrams are shown in Figure~\ref{fig:col_mag},
using the same notation, where filled circles denote GCs matched with
X-ray point sources.  Again, the bimodality of the GC system is
apparent in all color indices.  These color-magnitude plots, or the
color-color plots (Figure~\ref{fig:col_col}) indicate that the
GCs matched with X-ray sources, although redder, fall
in general among the bulk of the GC population.

A few outlying X-ray sources in the $BVI$ diagram are quite faint, and
could represent contaminant sources (foreground stars or background
galaxies).  Furthermore, a few objects, shown in the $(B-V)$ versus
$V$ plot, show slightly bluer colors than the bulk of the confirmed
population.  All 12 matched X-ray sources with $(B-V)\! < \!0.5$ are
unconfirmed GC candidates (see Table~2), and could therefore be 
younger star clusters, or simply contamination. 

There are only two confirmed clusters, neither matched to an X-ray
source, with $(B-V) \!<\! 0.5$. These GCs (GC0084 and GC0103) have
ages $\leq 1$ Gyr, as estimated by \cite{pff04II}. One of them
(GC0103) appears to be located in a young blue tidal stream in
the galaxy \citep{peng02}.  It is possible that all of the GC candidates
 with associated X-ray binaries and very blue colors
are young clusters, or perhaps even background galaxies with X-ray
emission.  Only spectroscopic follow-up can conclusively determine
their true nature.  The positions of these blue objects are clearly
seen in Fig.~\ref{fig:BV_rad}.  

\subsection{Dependence on Metallicity}

The $(B-I)$ color index, which is primarily sensitive to metallicity
for old stellar populations \citep[e.g.][]{worthey94}, has been
shown to clearly resolve the bimodal GC populations in
several brightest cluster galaxies.  In a study of eight galaxies,
\citet{har06} found that the blue and red cluster populations have
$\langle B-I\rangle_0 = 1.64\pm0.03$ and $\langle B-I\rangle_0 =
2.06\pm0.05$ respectively, and internal dispersions of $\sigma_{B-I} =
0.10\pm0.02$ and $\sigma_{B-I} = 0.17\pm0.05$, respectively.  Assuming
a foreground absorption of $E(B-I) = 0.273$ for Cen~A, the
transition between the metal-rich and metal-poor cluster populations
is roughly at $(B-I) \simeq 2.072$.  This division also agrees with
the uncorrected color distributions of \cite{pff04II}.   We have used 
[Fe/H] data for confirmed clusters and  $(B-I)$ colors when
no [Fe/H] data is available, to determine the fractions of
metal-rich to metal-poor clusters hosting X-ray sources.

There is an obvious trend in the metallicity distribution of the GCs
that contain X-ray point sources, when compared to the entire sample.
The entire list of confirmed GCs contains 179 ($53\%$)
metal-poor and 156 ($47\%$) metal-rich clusters of the 335
GCs with available metallicity data, closely matching
previous studies \citep{hhg04II,WHH05}.  But of the
confirmed GCs with X-ray sources, 6 ($21\%$) are metal-poor
and 23 ($79\%$) are metal-rich clusters out of the 29 confirmed GC matches with
available metallicity data.  Of the additional candidate
GC LMXBs, 16 ($52\%$) are metal-poor (blue) and 15 ($48\%$) are
metal-rich (red) objects from the 32 candidate GCs with available
metallicity data. From
a straight combination of confirmed and candidate 
GCs, we therefore find that 22 ($37\%$) of LMXBs match metal-poor GCs and 38
($63\%$) match metal-rich ones out of the 60 GCs with available data. 
We note though that some fraction of the GC candidates,
particularly those with $(B-V) \!<\! 0.5$ (see Figure~\ref{fig:col_mag}),
might be interlopers or very young clusters, so the fraction matched to 
the metal-rich clusters is probably a lower limit. We conclude therefore
that the ratio of metal-rich to metal-poor GCs with associated LMXBs
is $\gtrsim1.7$. 
While previous observations of early-type galaxies gives a ratio of
$\approx 3$, the scatter around this value is very large \citep{siv07} 
and comfortably includes our lower limit.
If we consider only the confirmed GCs, the ratio is 3, more in line
with the mean of the values found in previous studies.

\subsection{Dependence on GC Mass or GC Luminosity}

Finally, we explore the dependence of the incidence of point sources,
which are predominantly LMXBs, on the luminosities (and indirectly on
the masses) of the host GCs. Previous studies have found that X-ray point
sources are preferentially found in optically more luminous GCs
\citep[e.g.][]{angelini01,kundu03,sarazin03,jordan04,siv07}, 
while in the Galaxy and in M31 they have been noted to occur in 
the denser GCs \citep{bellazzini95}. 
Because of the fact that the sizes of GCs do not correlate with their
mass for masses $M \lesssim 2 \times 10^6 M_\odot$
\citep{mclaughlin00,jordan05} a preference of LMXBs 
in massive clusters translates into a preference for {\it denser} GCs, which
in the picture of dynamically formed LMXB progenitors is the
most fundamental property.
In the left panel of Figure~\ref{fig:dist-I}, we plot the
distribution of the $I$-band magnitudes of all the GCs (open
histogram) along with the magnitude distribution of those that match
point sources (filled histogram). As before, the hatched histogram
corresponds to all GCs out to 16.5~kpc. While LMXBs are hosted by 
GCs having luminosities in a wide range, it is clear that the distribution
of GCs that host an LMXB does not follow the parent GC distribution.
Instead, the distribution is much flatter as a consequence of a higher
fraction of massive GCs hosting LMXBs.

In the right-hand panel of Fig.~\ref{fig:dist-I}, in each one-magnitude
bin, we have calculated the total $I$-band luminosity of all GCs by
assuming the adopted distance of 3.8 Mpc (for
GCs with no $I$ magnitude, we use an estimate from their $T1$
magnitude).  We plot the number of matched X-ray point sources per unit
GC luminosity (for all GCs with $r<16.5$~kpc) in each bin. The errors
reflect Poisson errors in the number of GCs.  We have made an
approximate correction for completeness, by comparing the number of
GCs in each magnitude bin to those predicted by the GC luminosity
function and radial distribution of GCs in \cite{har06}.  This
correction is very large and uncertain fainter than $I=19.87$, 
which is indicated by the dashed vertical line.

The number of matched point sources
(or, equivalently,  the number of matched GCs)
{\it per unit GC luminosity} is consistent with being constant
across the luminosity range of our GC sample. We determine
that GCs in Cen~A host on average $\approx 0.4$ LMXBs per
$10^6 L_{\odot,I}$ (uncorrected luminosities). In other
words, and consistent with previous studies 
\citep[e.g.][]{sarazin03,jordan04,siv07},
we find that the number of X-ray point sources per unit GC
luminosity is roughly constant or, equivalently, that the probability
of a GC hosting an LMXB is roughly proportional to its mass.
\citet{sarazin03} found in their sample of four galaxies that GCs
host $\approx 0.15$ LMXBs per $10^6 L_{\odot,I}$. Their lower value,
compared to the one found here,
can be understood as the result of the fact that their LMXB samples 
were restricted to higher X-ray luminosities than ours (their
X-ray point source sample has mostly $L_X \gtrsim 10^{38}$ erg~s$^{-1}$).
Therefore, they probe a smaller fraction of the LMXB population which 
naturally implies a lower observed number of LMXBs per unit GC mass.

\section{Properties of the matched X-ray point Sources}
\label{s:xraypts} 

The four Chandra/ACIS observations explored in this work together cover
almost all (about 310 sq.arcmin) of the circle of radius 10 arcmin centered 
on the galaxy (see Fig.~2, also \citet{vg06}).  According to
the \citet{moretti03} compilation of 
point-source counts various ROSAT and Chandra Deep Field surveys, the number
of background sources (mostly AGN) expected down to a 0.5--2~keV flux limit
of  $f_X\!>\!1\times 10^{-15}$ erg~s$^{-1}$ cm$^{-2}$
and $f_X\!>\!5\times 10^{-16}$ erg~s$^{-1}$ cm$^{-2}$ within
this area of the sky is 67 and 104 respectively. We find 153 and
185 X-ray point sources brighter than these flux limits 
respectively within this circle.  Here we
use the 0.5--2~keV flux for each point source, taking the mean
value for sources that are detected in multiple observations.

If we take a smaller circle, that of radius 5~arcmin, from the centre,
the background counts down to the flux limit of
$f_X\!>\!1\times 10^{-15}$ erg~s$^{-1}$ cm$^{-2}$
and $f_X\!>\!5\times 10^{-16}$ erg~s$^{-1}$ cm$^{-2}$ within
this radius are 17 and 26, whereas  we detect 92 and 114 point sources
down to these flux limits respectively. 

In Figure~\ref{fig:logn-logs}a, we plot the luminosity function (LF) of
the matched X-ray point sources (solid histogram) within 5\arcmin\ of
the center of Cen~A.  There are 36 matched
point sources within this region.  Of these, 30 have mean 0.5--2~keV 
fluxes more than
$f_X\!>\!5\times 10^{-16}$ erg~s$^{-1}$ cm$^{-2}$. 
Down to this flux limit, there are
84 sources that are not matched with any confirmed or 
candidate GC. The unmatched sources are 
represented by the red unbroken histogram in the same plot.

\begin{figure}
\includegraphics[width=0.9\hsize]{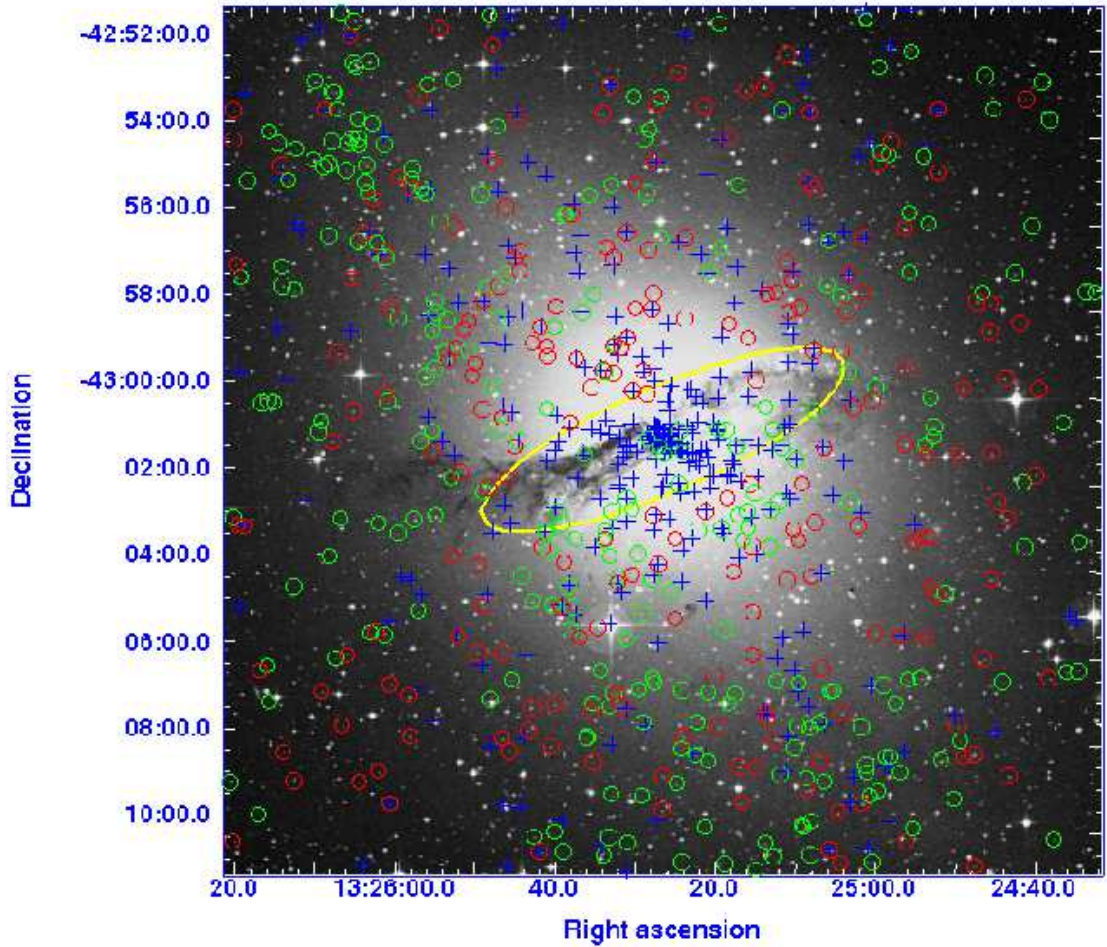}
\caption{A optical image of the early-type galaxy Centaurus~A
  (NGC~5128), together with
  the X-ray point sources
  (blue crosses), the confirmed GC population (red open circles)
  and the GC candidates unconfirmed by spectroscopy
  (green open
  circles) in the inner region of Cen~A. The optical image is from
  the IIIa-J blue UKSTU plates, obtained
  from the STSCI Digitized Sky Survey (http://archive.stsci.edu/).}

\label{fig:positions}
\end{figure}

\begin{figure}
\includegraphics[width=0.5\hsize]{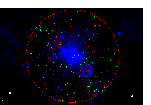}
\includegraphics[width=0.5\hsize]{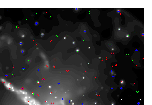}
\caption{(a) A soft X-ray image (0.5-2.0 keV) of 
  NGC~5128, adaptively smoothed and exposure corrected combined image
  with the confirmed GC population (red open circles)
  and the GC candidates unconfirmed by spectroscopy
  (green open
  circles), within 10 arcmin of the center of the galaxy
  (marked as a circle). (b) A close-up of a 
part of the galaxy, where the detected X-ray point
  sources above the flux limit of this paper are shown as 
blue crosses as well, with a GCs shown as circles as before.}

\label{fig:x-positions}
\end{figure}

\begin{figure}
\centering
\includegraphics[width=0.6\hsize]{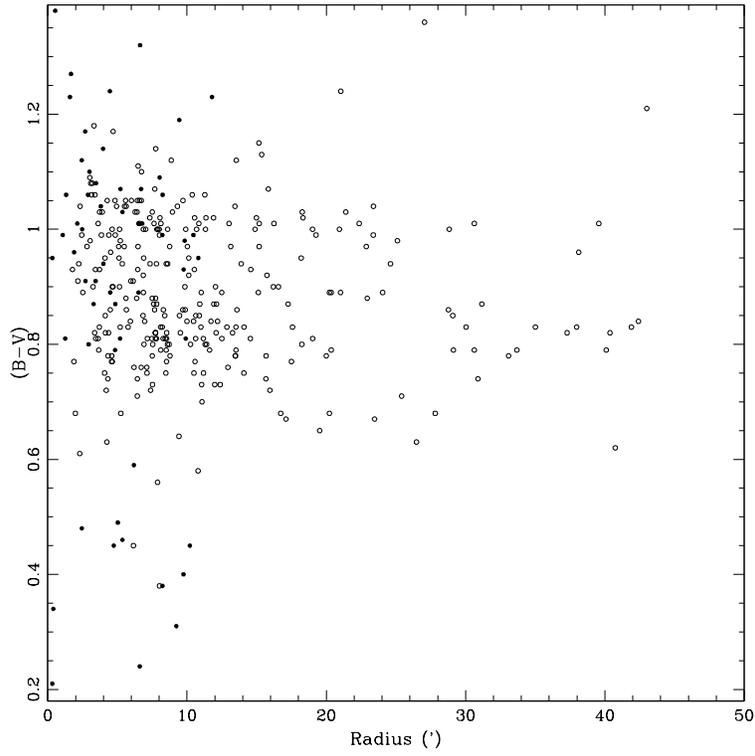}
\caption{The radial distribution of the confirmed sample of 
  GCs in Cen~A (open circles) and the GCs and
  candidates with X-ray matches (closed circles) in $(B-V)$.  
  GC candidates with X-ray matches with a color of $(B-V) \!<\!
 0.5$ are too blue to be old GCs.}
\label{fig:BV_rad}
\end{figure}

\begin{figure}
\centering
\includegraphics[width=0.6\hsize]{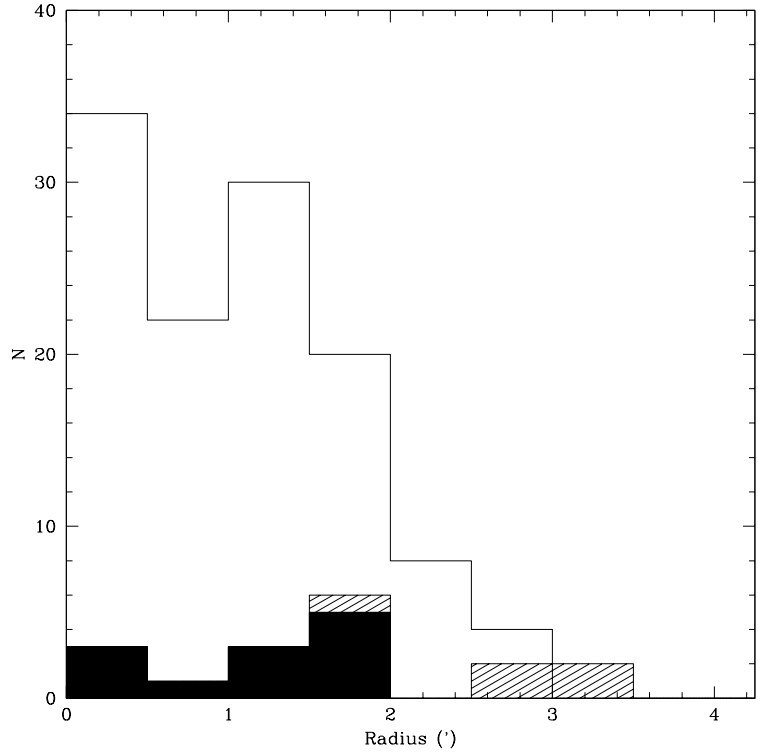}
\caption{The radial distribution of X-ray point sources and
GCs in the inner 4\minpoint 2 occupied by the Cen~A dust lane.
The open histogram is for all X-ray point sources, the hatched
histogram is known GCs, and the solid histogram is all GC 
X-ray sources enclosed in the exclusion ellipse (see text).} 
\label{fig:dustlane}
\end{figure}

\begin{figure}
\centering
\includegraphics[width=0.9\hsize]{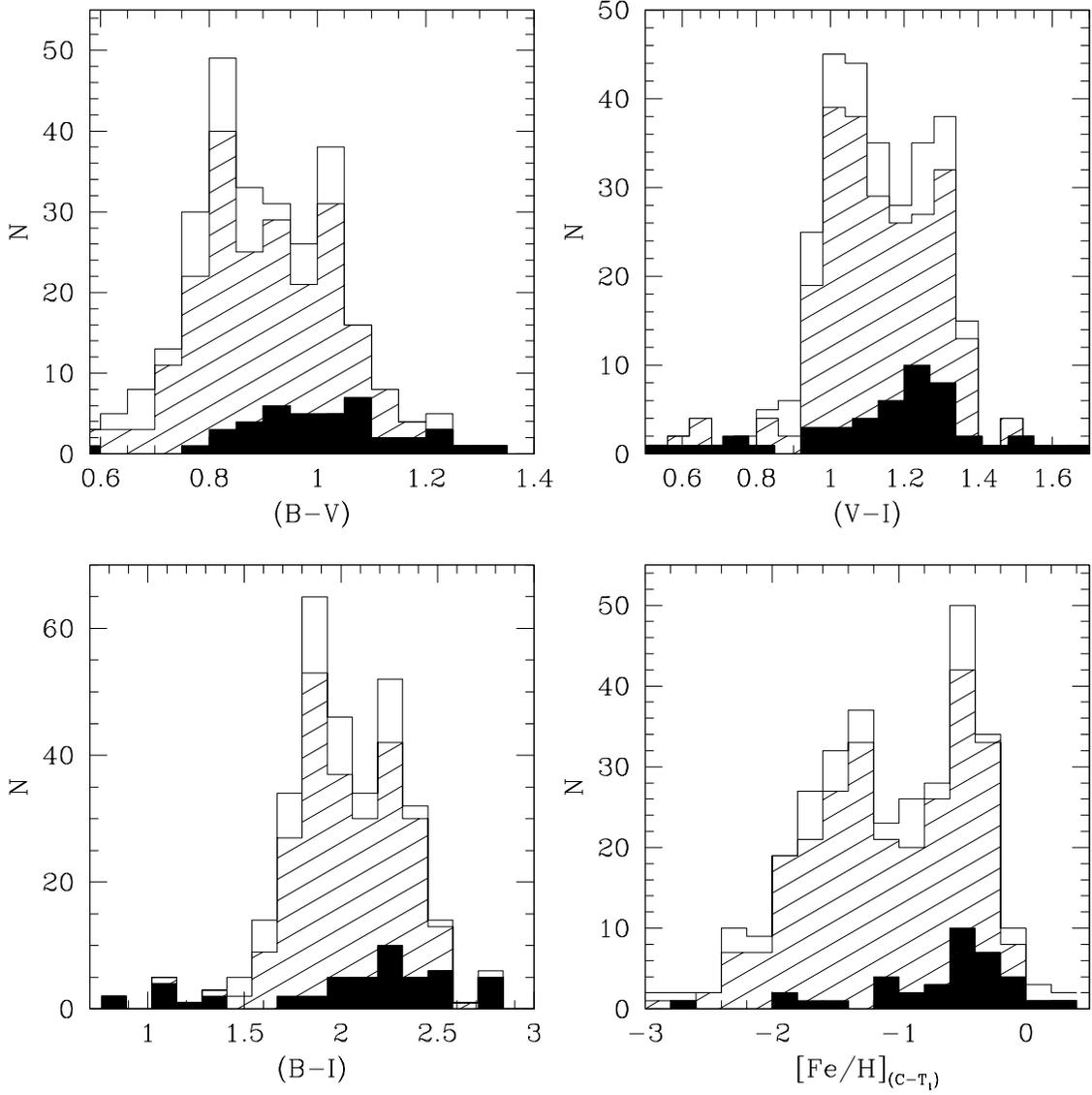}
\caption{The color distributions  and the metallicity distribution
function 
of the entire sample of
  GCs in Cen~A (open histograms), and the GCs
  matched with an X-ray source (solid histograms) in ({\it upper left})
  $(B-V)$, ({\it upper right}) $(V-I)$, ({\it lower left}) $(B-I)$, and
  ({\it lower right}) $[Fe/H]_{(C-T_1)}$. The hatched histograms
  represent
  all GCs within 16.5 kpc of the center of the galaxy, the radius out
  to which LMXBs are detected in this study.
} 
\label{fig:distributions}
\end{figure}

\begin{figure}
\centering
\includegraphics[width=0.9\hsize]{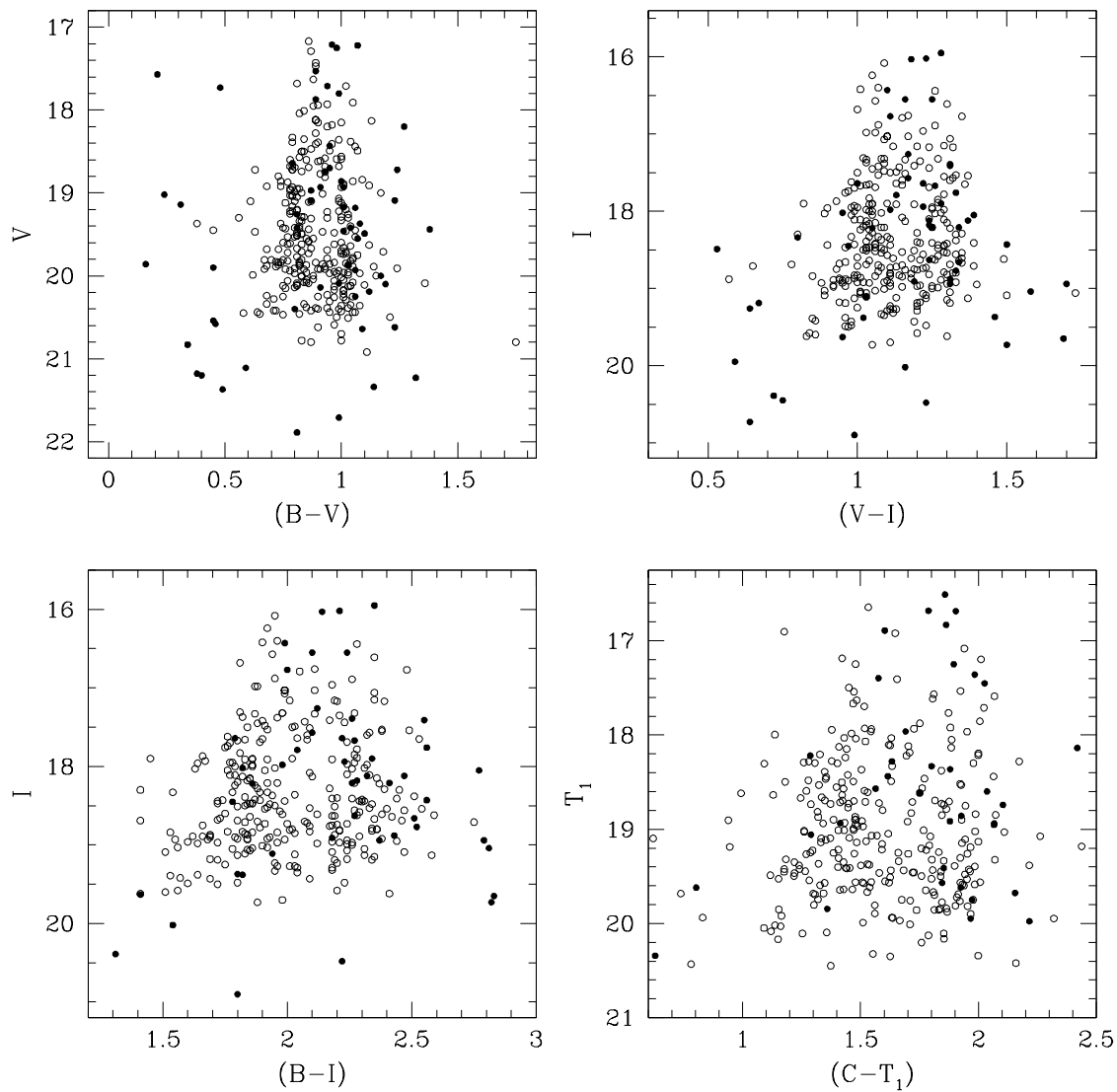}
\caption{The color-magnitude diagrams of the sample
  of GCs in Cen~A (open circles) and the
  GCs and candidates with X-ray matches (closed circles)
in ({\it upper
  left}) $(B-V)$ versus $V$ magnitude, ({\it upper
  right}) $(V-I)$ versus $I$ magnitude, ({\it lower left}) $(B-I)$
  versus $I$ magnitude, and ({\it lower right})
  $(C-T_1)$ versus $T_1$ magnitude.} 
\label{fig:col_mag}
\end{figure}

\begin{figure}
\centering
\includegraphics[width=0.9\hsize]{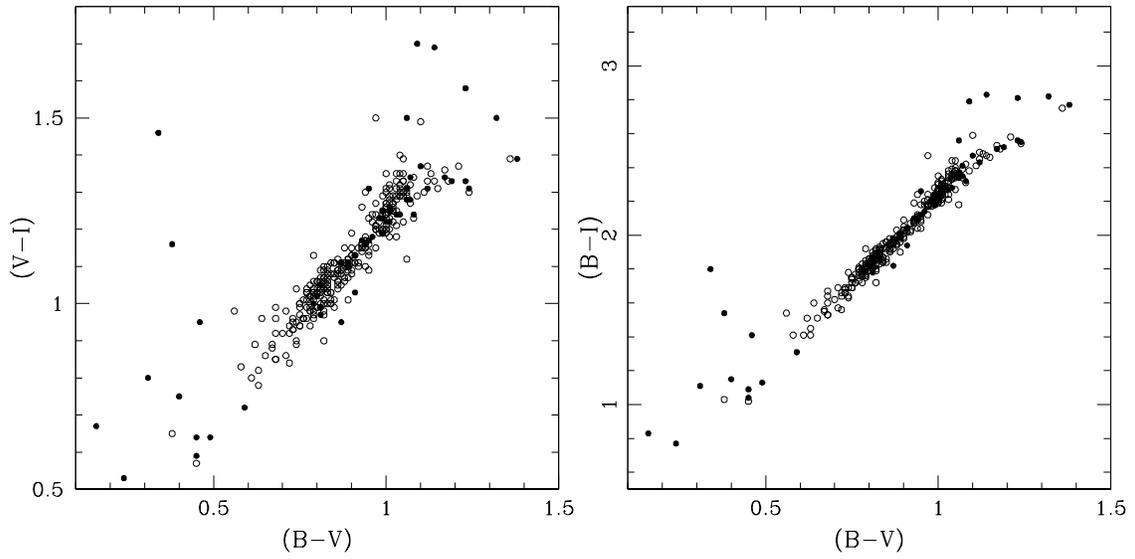}
\caption{The color-color diagrams of the sample of 
GCs in Cen~A (open circles) and the GCs and
  candidates with X-ray matches, in ({\it left}) $(B-V)$ versus $(V-I)$
  and ({\it right}) $(B-V)$ versus $(B-I)$.} 
\label{fig:col_col}
\end{figure}

\begin{figure}
\centering
\includegraphics[width=0.45\hsize]{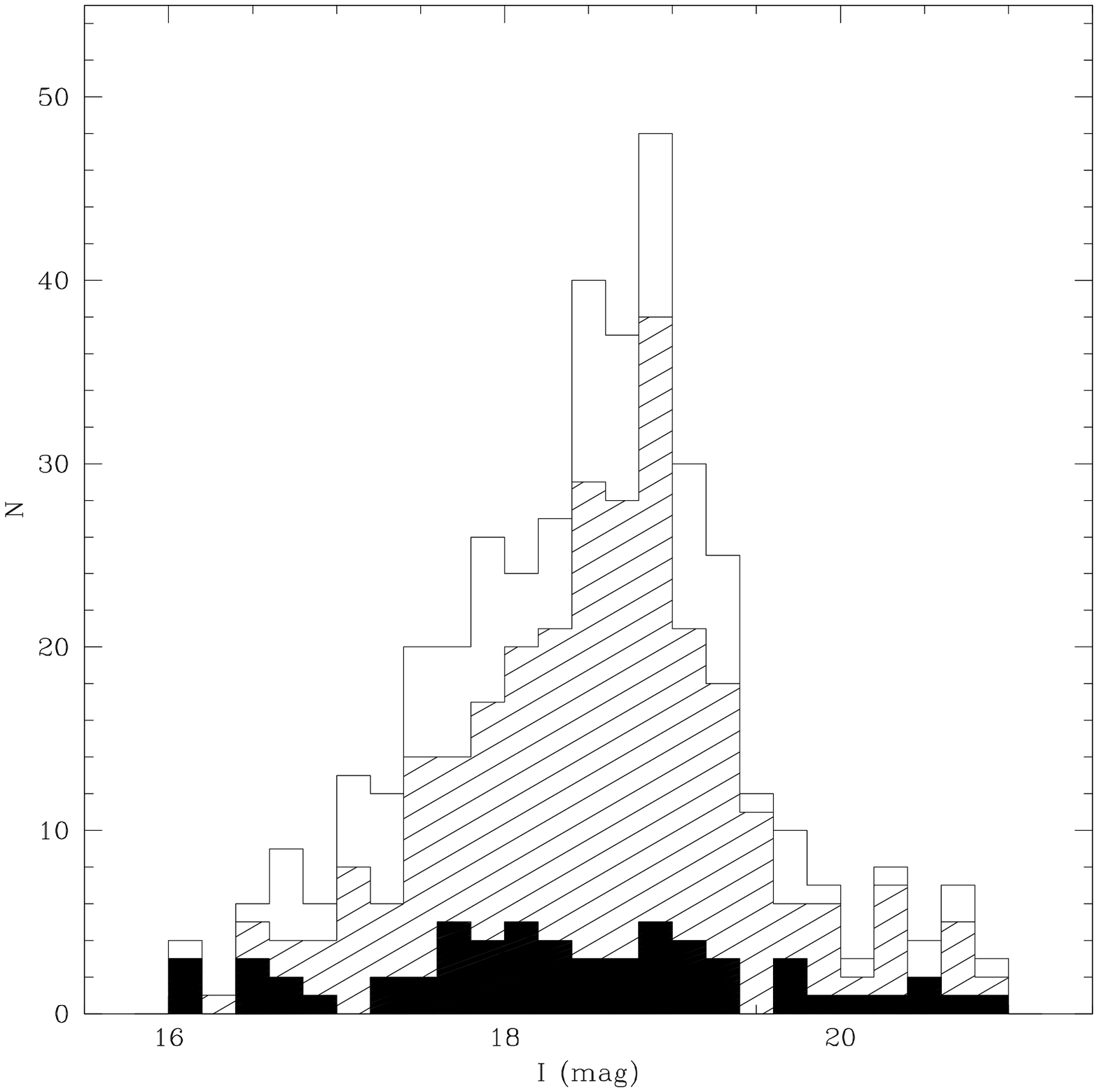}
\includegraphics[width=0.45\hsize]{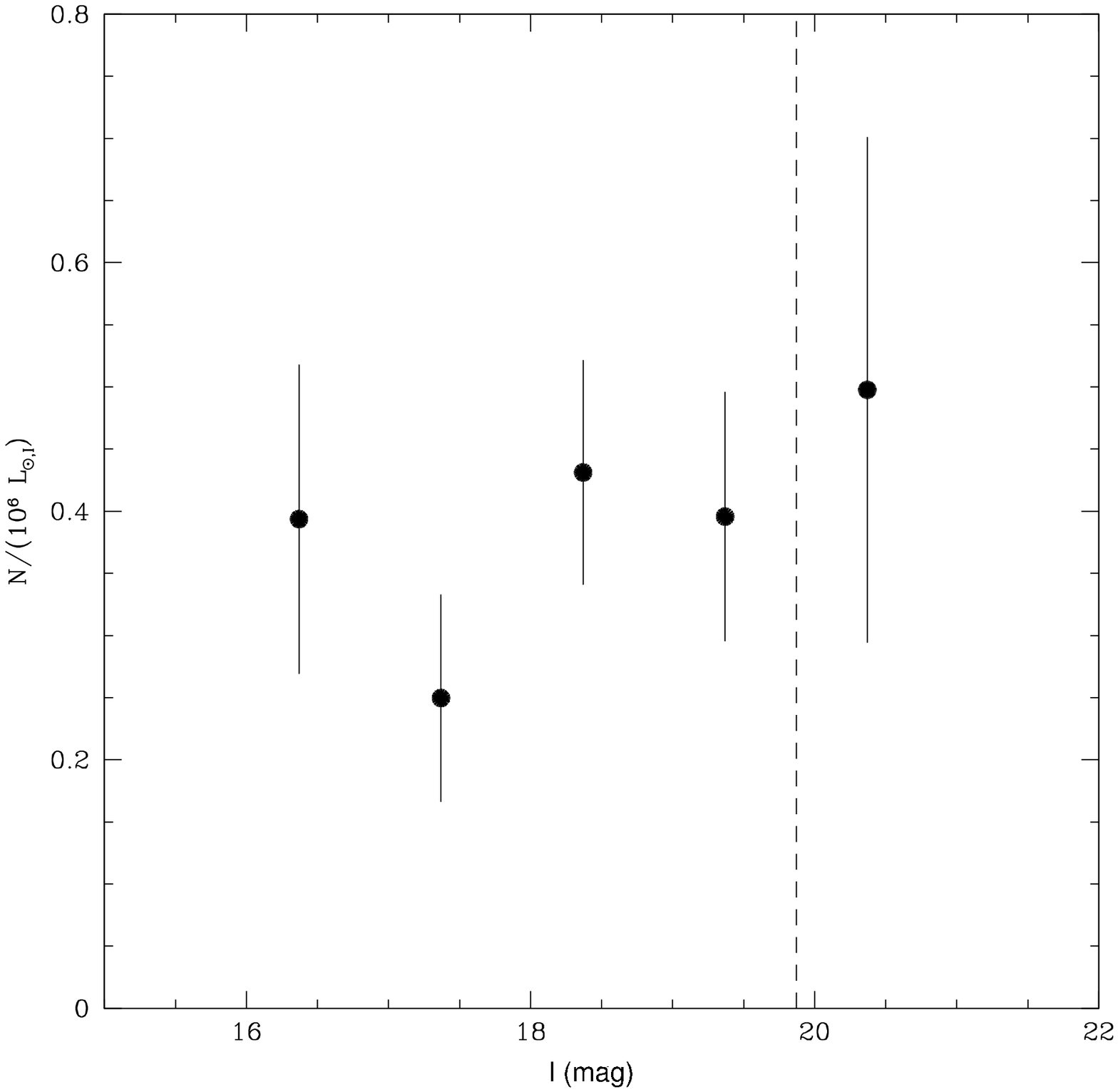}
\caption{(a) The distribution of total apparent $I$ magnitude of the
  GCs used in this study, where the open histogram represents
all  confirmed GCs in Cen~A, and the solid histogram shows GCs
  matched with an X-ray source. The hatched histogram represents
  all GCs within 16.5 kpc of the center of the galaxy, the radius out
  to which LMXBs are detected in this study.
There is no obvious trend of finding LMXBs with GC luminosity. (b) 
The number of matched X-ray point sources per unit
GC luminosity (for all GCs with $r<16.5$~kpc) as a function
of $I$ magnitude of the GCs.  In each 
one-magnitude bin,
the total $I$-band luminosity of all GCs is calculated by
assuming them to be all at the adopted distance of the galaxy, 
and an estimated correction is made for incompleteness 
(see \S\ref{s:gcmatches}).  The number of matched LMXBs per unit light is
  independent of GC luminosity.
The sample
is seriously incomplete fainter than $I=19.87$, indicated by the
dashed vertical line.
}
\label{fig:dist-I}
\end{figure}

\begin{figure}
\centering
\includegraphics[angle=-90,width=0.45\hsize]{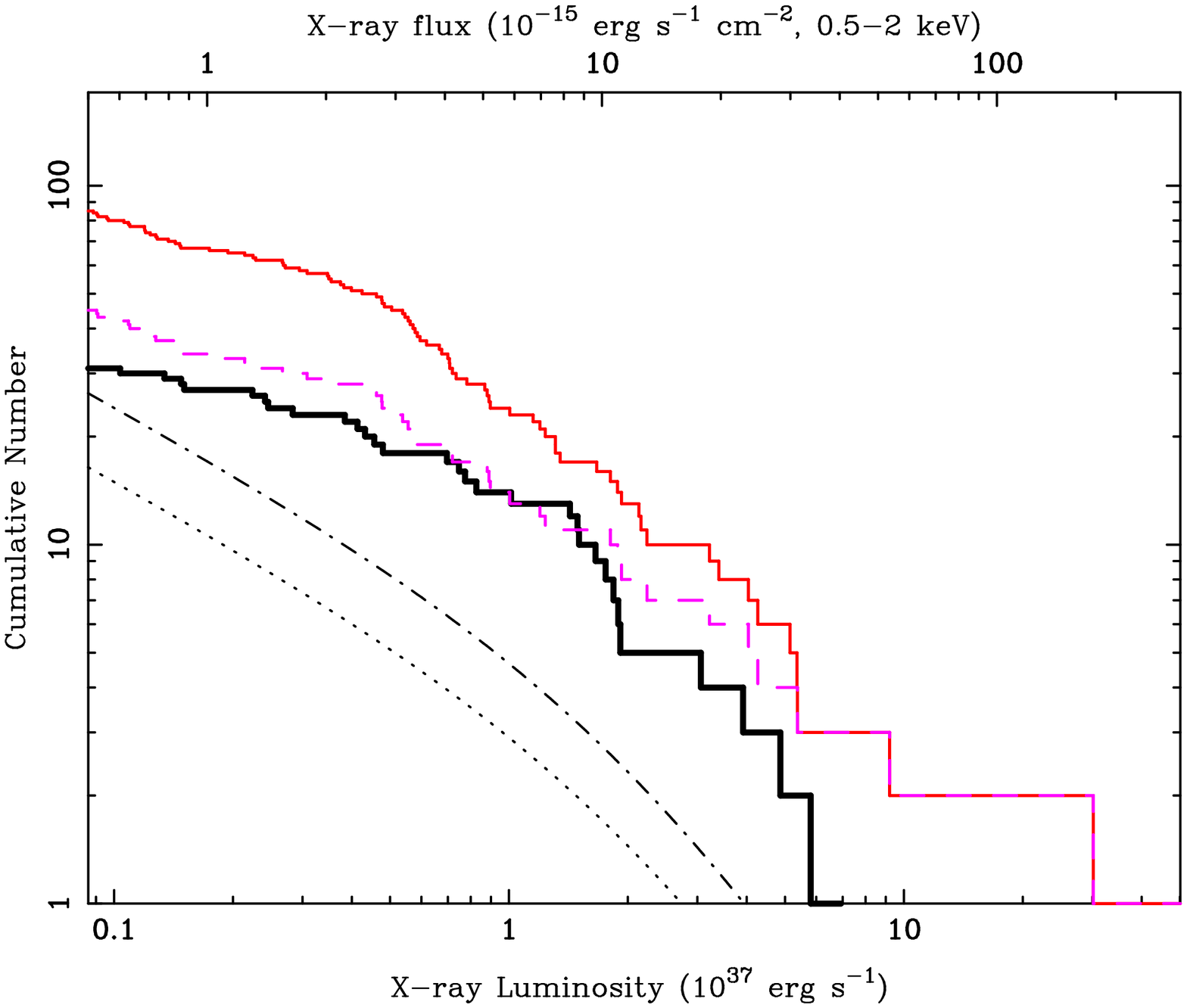}
\includegraphics[angle=-90,width=0.45\hsize]{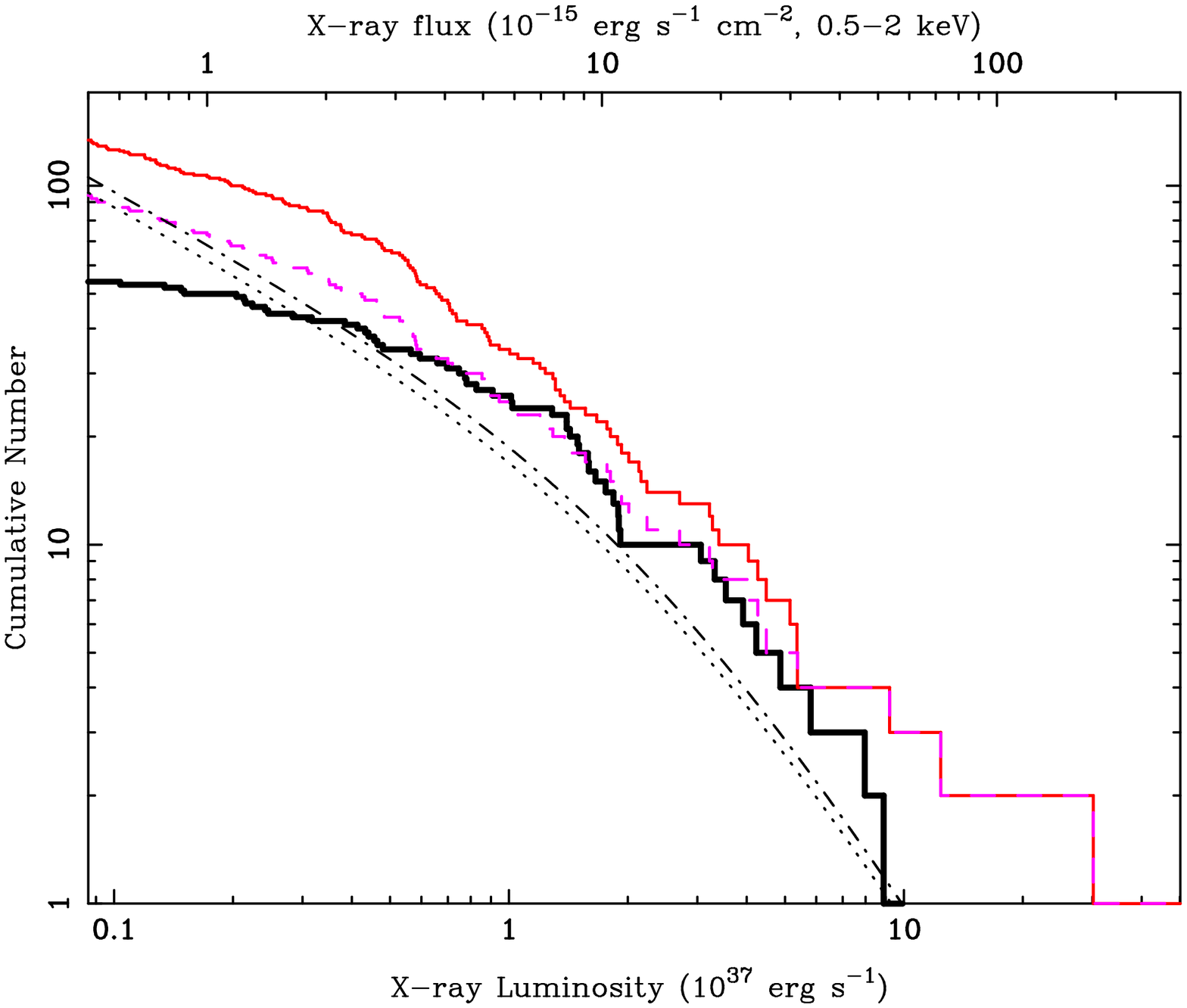}
\caption{
(Left:) The luminosity function (LF) of all 30 X-ray point sources,
brighter than $5\times 10^{-16}$ erg~s$^{-1}$ (0.5--2 keV), found within
5\arcmin\ of the center of Cen~A (solid histogram), that are found to
match a GC. For comparison, we also plot the LF of all the X-ray point
sources that did not match GCs (red solid histogram, $N\!=\! 84$).  Of
these, the unmatched source within 5\arcmin, but outside the dust lane
region ($N\!=\!44$ brighter than $5\times 10^{-16}$ erg~s$^{-1}$)
are shown as the dashed purple histogram.  The
black dash-dotted and dotted curves represent the expected counts of
backgound X-ray sources (mostly AGN and background galaxies) estimated
from deep blank sky surveys, compiled by \citep{moretti03}, normalized
to the area covered by the solid red histogram and dashed purple
histogram respctively. Clearly the background sources do not account
for almost half of the unmatched sources above a flux of
$\sim 5\times 10^{-16}$ erg~s$^{-1}$ cm$^{-2}$, even outside the dust
lane region where the GC catalogue is expected to be reasonably
complete. In this plot, we use the 0.5--2~keV luminosity of each point
source, taking the average value for sources that are detected in
multiple observations. 
(Right:) The same curves, but for all sources
found within 10\arcmin\ of the center of Cen~A. Outside the 5~arcmin
region, the fraction of unmatched sources accounted for by background
sources is subtantially larger.
\label{fig:logn-logs}
}
\end{figure}

It can be argued that a significant fraction
of the unmatched X-ray sources may lie in the dust lane region, 
where there are far fewer detected GCs than X-ray point sources
(see Fig.~\ref{fig:positions}) .
To investigate this, we define the dust lane region to be an
ellipse with a
semi-major axis of 4.5\arcmin, parallel to the dust lane, and a
semi-minor axis of 1.2\arcmin, thus 
occupying 22\% (but in the core of the galaxy)
of the surveyed area within the 5\arcmin\ circle
(see Fig.~\ref{fig:positions}).
Outside this dust
lane region, shown as the purple dashed curve, 44
of the 84 unmatched sources with
flux $f_X\!>\!5\times 10^{-16}$ erg~s$^{-1}$ cm$^{-2}$
exist.
This shows that if one excludes the dust lane region, the 
numbers of matched and unmatched point sources
are comparable, down to a luminosity of about $L_X\!=\!8.6\times
10^{36}$ erg~s$^{-1}$.
For the distributions of these 30 and 44 objects respectively,
a Kolmogorov-Smirnov (K-S)
test shows that the probability that the two subsamples could be drawn
from the same population is about 54\%. 

The black dash-dotted and dotted curves represent the expected counts
of background X-ray sources (mostly AGN and background galaxies)
estimated from the \citet{moretti03} compilation, normalized to the
area covered by the solid red histogram and dashed purple histograms
respctively.  Brighter than $\sim 5\times 10^{-16}$ erg~s$^{-1}$
cm$^{-2}$, there are 26 background sources
expected within the 5\arcmin\ circle, but only 16 if one considers the
area within the circle excluding the dust lane.  So outside the dust
lane, within 5\arcmin, 36\% (=16/44) of the unmatched sources are
expected to be background ones.  
It must also be noted that \citet{vg06} find that in the Cen~A field, the
density of X-ray point sources is significantly higher than
the average values found by studies by that of 
\citet{moretti03}. 
Even with the enhanced background counts,
the background
point source population does not account for at least half of
the unmatched sources above a flux of $\sim 5\times 10^{-16}$
erg~s$^{-1}$ cm$^{-2}$, even outside the dust lane region, where the
GC catalogue is expected to be reasonably complete.

In Figure~\ref{fig:logn-logs}b, we plot the 
same curves for
the X-ray point sources  within 10\arcmin\ of
the center of Cen~A.  There are 53 matched
point sources within this region with
flux more than
$f_X\!>\!5\times 10^{-16}$ erg~s$^{-1}$ cm$^{-2}$, 
whereas there are
133 sources that are not matched with any confirmed or 
candidate GC, 93 of which are outside the dust lane region.
Unlike the above case of $<$5~arcmin objects, the subsamples
of the matched 53 sources and unmatched 93 sources
have a 1.7\% chance of being drawn from the same population,
according to the K-S test. This shows that outside the 5\arcmin\
circle, the detected point sources are increasingly 
not associated with the galaxy. Indeed,
of these 93 unmatched sources out to 10~arcmin, 
62 are expected to be background sources
from the \citet{moretti03} counts, and higher according
to the \citet{vg06} study.

\section{Concluding remarks}
\label{s:remarks} 

We have presented a study of the connection between GCs and LMXBs
using a ground-based catalog of GCs and an X-ray point source list
constructed using four Chandra/ACIS
X-ray observations of the nearby early-type galaxy Cen~A, obtained
over a four-year period, each with an exposure time of 37--50~ks.  Of
the 353 X-ray point sources detected on one or more of these Chandra fields,
67 are found to be hosted by known GCs or 
GC candidates. Thirty of these matches correspond to
GCs confirmed by radial velocity, and 1 match corresponds to a
resolved GC by HST imaging. 
Even in the central heavily obscured dust lane, we find matches between the two
populations, although the optical samples are significantly incomplete
in that region. 

As in many other early-type galaxies \citep[e.g. ][]{peng06}, 
the GCs in Cen~A have a large
range in metallicity and a bimodal color distribution.  A three-to-one
majority of the matched X-ray point sources are seen to lie in the
redder, more metal-rich GCs: specifically, 156/335 
($44\%$) of the confirmed GCs are in  the metal-rich subpopulation, 
while 23/29 ($79\%$) of the X-ray point sources matched to confirmed
GCs lie in these clusters.  We also note a possible double X-ray
point source detection in the radially velocity confirmed GC0233.

Also in agreement with previous studies, we find that luminous GCs
are most likely to host LMXBs. Quantitatively, the number of LMXBs
per unit GC mass is roughly constant.  The lack of correlation
between GC size and mass implies that denser GCs are more likely
to host LMXBs, as expected if most GC LMXB are formed dynamically by
tidal capture or exchange processes in GC cores.

Our expanded sample of GC-LMXB identifications in Cen~A gives
further support for a consistent picture which has been emerging
from combined optical and {\it Chandra} observations of early-type
galaxies. Important factors determining the presence of LMXBs in GCs
which can be easily determined from ground-based observations
are their mass, which is a reflection of a more fundamental
dependence on dynamical formation rates \citep{jordan07},
and their metallicity. The latter is a bona fide effect which seems 
not to follow from any possible trends of GC structural properties or 
age with metallicity.

The X-ray luminosity function of the LMXBs that are found in 
GCs and of those that are not matched with GCs are found to have 
the same slope, revealing similar underlying
populations. In fact, if one samples the areas of the galaxy 
that have more complete coverage for both the LMXB and GC populations
(away from the dust lanes and within 10\arcmin\ of the galaxy center),
half of the LMXBs are found within GCs.

However, there is some indication that, towards the faint
end of the luminosity function, the fraction of LMXBs 
found in GCs is far lower than at the more luminous end, but it is not
clear whether this is due to incompleteness effects at the faint end.
Cen~A offers a unique opportunity to study this question
because of its proximity, and we will probe the LMXB/GC connection
with our upcoming deep observations to X-ray luminosities unattainable
in more distant early-type galaxies.
In addition, we plan to use high-resolution optical
imaging material, from both ground based telescopes and from the HST/ACS,
to probe the structural properties of
the GCs with and without LMXBs, as well as the {\sl Chandra} spectral
properties of the X-ray sources.

\acknowledgments

This work was supported by NASA contracts NAS8-39073 and NAS8-03060,
the Chandra Science Center and the Smithsonian Institution.  KAW and
WEH acknowledge financial support from the Natural Sciences and
Engineering Research Council of Canada.



\clearpage

\begin{deluxetable}{lccc}
\tablecolumns{4}
\tabletypesize{\small}
\tablecaption{{Summary of {\em Chandra}/ACIS observations of
Centaurus A used here}\label{cxoobslog}}
\tablewidth{0pt}
\tablehead{
\colhead{OBSID} & \colhead{Date} & 
\colhead{Instrument} & \colhead{Cleaned Exposure Time} 
}\startdata
00316 &  5 Dec 99 & ACIS-I & 35724 s \\ 
00962 & 17 May 00 & ACIS-I & 36505 s \\ 
02978 &  3 Sep 02 & ACIS-S & 44592 s \\ 
03965 & 14 Sep 03 & ACIS-S & 49517 s \\ 
\enddata
\end{deluxetable}

\clearpage
\begin{deluxetable}{lccrcccccl}
\rotate
\tablecolumns{10}
\tablecaption{Globular Clusters in Cen~A (NGC 5128) with X-ray Sources
\label{tab:GCmatches}}
\tablewidth{0pt}
\tablehead{
\colhead{Cluster\tablenotemark{a}} & 
\colhead{RA} & \colhead{Dec} & \colhead{Radius} &
\colhead{$B$} &\colhead{$V$} &\colhead{$I$} &\colhead{$T_1$} &
\colhead{$(C-T_1)$}&\colhead{[Fe/H]}  \\
\colhead{ } &\colhead{(J2000)} & \colhead{(J2000)}
&\colhead{(arcmin)}& \colhead{(mag)}
 &\colhead{(mag)} &\colhead{(mag)} &\colhead{(mag)} &\colhead{ } &\colhead{ }\\
}\startdata
GC0074\tablenotemark{b}  &13 24 54.35 &-42 53 24.8&  9.84& 18.24& 17.25& 16.02& 16.68&  1.788&  -0.6732\\
GC0109\tablenotemark{b}  &13 25 03.13 &-42 56 25.1&  6.51& 19.94& 18.93& 17.67& 18.36&  1.880&  -0.5088\\
GC0120\tablenotemark{b}  &13 25 05.02 &-42 57 15.0&  5.68&   -  &   -  &   -  & 16.83&  1.863&  -0.5379\\
GC0123\tablenotemark{b}  &13 25 05.72 &-43 10 30.7& 10.18&   -  &   -  &   -  & 17.36&  1.984&  -0.3429\\
GC0129\tablenotemark{b}  &13 25 07.62 &-43 01 15.2&  3.66&   -  &   -  &   -  & 17.45&  2.026&  -0.2818\\
GC0134\tablenotemark{b}  &13 25 09.19 &-42 58 59.2&  4.00& 18.65& 17.71& 16.55&   -  &    -  &     -  \\
GC0137\tablenotemark{b}  &13 25 10.25 &-42 55 09.5&  6.78& 20.47& 19.46& 18.21& 18.86&  1.927&  -0.4248\\
GC0138\tablenotemark{b}  &13 25 10.27 &-42 53 33.1&  8.23& 18.79& 17.80& 16.55& 17.25&  1.895&  -0.4769\\
GC0141                   &13 25 11.05 &-43 01 32.3&  3.05&    - &   -  &   -  &   -  &    -  &     -   \\	
GC0158\tablenotemark{b}  &13 25 14.24 &-43 07 23.5&  6.71& 20.63& 19.55& 18.21& 18.95&  2.066&  -0.2269\\
GC0191\tablenotemark{b}  &13 25 22.19 &-43 02 45.6&  1.90& 18.17& 17.21& 16.03& 16.51&  1.858&  -0.5466\\
GC0205\tablenotemark{b}  &13 25 27.97 &-43 04 02.2&  2.89& 20.23& 19.18& 17.90& 18.60&  2.036&  -0.2678\\
GC0210\tablenotemark{b}  &13 25 29.43 &-42 58 09.9&  3.00& 20.58& 19.49& 18.12& 18.94&  2.067&  -0.2257\\
GC0217\tablenotemark{b}  &13 25 30.29 &-42 59 34.8&  1.64&   -  &   -  &   -  &   -  &    -  &     -  \\
GC0225\tablenotemark{b}  &13 25 31.60 &-43 00 02.8&  1.32& 20.99& 19.93& 18.43&   -  &    -  &     -  \\
GC0230\tablenotemark{b}  &13 25 32.42 &-42 58 50.2&  2.47& 19.86& 18.86& 17.64& 18.33&  1.801&  -0.6489\\
GC0231\tablenotemark{b}  &13 25 32.80 &-42 56 24.4&  4.83& 19.43& 18.64& 17.64& 18.22&  1.287&  -1.8618\\ 
GC0233\tablenotemark{b,c}&13 25 32.88 &-43 04 29.2&  3.48& 20.44& 19.37& 18.12& 18.74&  2.104&  -0.1727\\
GC0246\tablenotemark{b}  &13 25 35.16 &-42 53 01.0&  8.25& 21.31& 20.25& 18.94& 19.75&  1.973&  -0.3534\\
GC0249\tablenotemark{b}  &13 25 35.50 &-42 59 35.2&  2.12& 20.18& 19.17& 17.94&   -  &    -  &      -  \\
GC0260\tablenotemark{b}  &13 25 38.61 &-42 59 19.6&  2.71& 19.83& 18.93& 17.79& 18.44&  1.616&  -1.0249\\
GC0266\tablenotemark{b}  &13 25 39.88 &-43 05 01.9&  4.49& 18.42& 17.53& 16.43& 16.89&  1.603&  -1.0539\\
GC0275\tablenotemark{b}  &13 25 42.00 &-43 10 42.2&  9.91& 20.08& 19.26& 18.22& 18.94&  1.414&  -1.5015\\
GC0281\tablenotemark{b}  &13 25 43.23 &-42 58 37.4&  3.81& 20.46& 19.42& 18.18& 18.92&  1.879&  -0.5105\\
GC0295\tablenotemark{b}  &13 25 46.59 &-42 57 03.0&  5.37& 20.89& 19.87& 18.63& 19.41&  1.853&  -0.5553\\
GC0314\tablenotemark{b}  &13 25 50.40 &-42 58 02.3&  5.20& 20.23& 19.42& 18.45& 19.06&  1.290&  -1.8476\\
GC0320\tablenotemark{b}  &13 25 52.74 &-43 05 46.4&  6.52& 18.76& 17.87& 16.77& 17.40&  1.576&  -1.1150\\
GC0330\tablenotemark{b}  &13 25 54.58 &-42 59 25.4&  5.22& 18.29& 17.22& 15.95& 16.69&  1.904&  -0.4686\\
GC0347\tablenotemark{b}  &13 26 00.81 &-43 09 40.1& 10.46& 21.09& 20.09& 18.91& 19.57&  1.846&  -0.5676\\
GC0378\tablenotemark{b}  &13 26 10.58 &-42 53 42.7& 10.81& 19.38& 18.43& 17.26& 17.96&  1.691&  -0.8644\\
GC0384\tablenotemark{b}  &13 26 19.66 &-43 03 18.6&  9.76& 19.68& 18.74& 17.57& 18.28&  1.633&  -0.9876\\
BTC136                   &13 25 01.16 &-43 11 59.0& 11.87&   -  &   -  &   -  & 19.68&  2.155&  -0.1161\\
BTC317                   &13 25 45.63 &-43 01 15.5&  3.30& 19.84& 18.97& 18.02&    - &    -  &     -   \\
Minniti03                &13 25 38.18 &-42 58 15.3&  3.48&   -  &   -  &   -  &    - &    -  &     -   \\
Minniti05                &13 25 33.95 &-42 58 59.5&  2.45& 18.21& 17.73&   -  & 18.62&  1.751&  -0.7440\\
Minniti17                &13 25 25.76 &-43 00 56.0&  0.40& 21.17& 20.83& 19.37&    - &    -  &     -   \\
Minniti23                &13 25 20.86 &-43 00 53.7&  1.26& 22.70& 21.89& 20.90&    - &    -  &     -   \\
Minniti26                &13 25 42.13 &-43 03 19.7&  3.44& 21.05& 20.14& 19.11&    - &    -  &     -   \\
Minniti33                &13 25 37.46 &-43 01 31.4&  1.84& 19.12&   -  &   -  &    - &    -  &     -   \\
Minniti40                &13 25 32.02 &-43 02 31.6&  1.60& 20.32& 19.09& 17.76&    - &    -  &     -   \\
Minniti44                &13 25 29.45 &-43 01 08.4&  0.34& 19.65& 18.70& 17.39&    - &    -  &     -   \\
Minniti53\tablenotemark{d}&13 25 27.49&-43 01 28.4&  0.33& 17.78& 17.57&    - &    - &    -  &     -   \\
Minniti55                &13 25 25.59 &-43 02 09.7&  1.08& 22.70& 21.71& 20.48&    - &    -  &     -   \\
Minniti57                &13 25 25.15 &-43 01 26.8&  0.54& 20.82& 19.44& 18.05&    - &    -  &     -   \\
Minniti71                &13 25 20.09 &-43 03 10.1&  2.45& 21.31& 20.19& 18.88& 19.62&  1.925&  -0.4344\\
Minniti74                &13 25 18.50 &-43 01 16.3&  1.67& 19.47& 18.20&   -  &    - &    -  &     -   \\
Minniti80                &13 25 14.03 &-43 02 42.9&  2.94&   -  &   -  &   -  & 19.85&  1.359&  -1.6584\\
Minniti82                &13 25 12.89 &-43 01 14.7&  2.69& 21.17& 20.00& 18.66&    - &    -  &     -   \\
Minniti208               &13 25 22.36 &-42 57 17.1&  3.98&   -  &   -  &   -  & 16.10& 3.251 &  -0.0206\\
Minniti216               &13 25 07.65 &-42 56 30.2&  5.91&   -  &   -  &   -  &    - &    -  &     -   \\
K-007                    &13 24 58.17 &-43 09 49.2& 10.21& 20.99& 20.54& 19.95& 20.34& 0.629 &  -4.1622\\
K-041                    &13 25 11.98 &-42 57 13.3&  4.86& 19.96& 19.09& 17.98& 18.57& 1.564 &  -1.1427\\
K-047                    &13 25 13.82 &-42 53 31.1&  8.04& 21.73& 20.64& 18.94& 19.95& 1.967 &  -0.3686\\
K-073                    &13 25 22.67 &-42 55 01.6&  6.19& 21.70& 21.11& 20.39& 20.64& 0.423 &  -5.0574\\
K-110                    &13 25 29.10 &-43 07 46.2&  6.63& 22.55& 21.23& 19.73& 20.44& 2.664 &   0.2205\\
K-116                    &13 25 31.05 &-43 11 07.1&  9.99& 20.02& 19.86& 19.19& 19.56& 0.498 &  -4.7219\\
K-141                    &13 25 34.05 &-43 10 31.1&  9.45& 21.29& 20.10& 18.77& 19.67& 0.000 &     -   \\
K-159                    &13 25 39.08 &-42 56 53.6&  4.74& 20.35& 19.90& 19.26& 19.62& 0.803 &  -3.4708\\
K-182                    &13 25 46.34 &-43 03 10.2&  3.98& 22.48& 21.34& 19.65&   -  & 0.000 &     -   \\
K-192                    &13 25 48.71 &-43 03 23.4&  4.46& 19.96& 18.72& 17.41& 18.14& 2.419 &   0.1217\\
K-201                    &13 25 53.75 &-43 11 55.6& 11.79& 21.85& 20.62& 19.04& 19.98& 2.216 &  -0.0490\\
K-204                    &13 25 55.13 &-43 01 18.3&  5.03& 21.86& 21.37& 20.73& 21.22& 1.024 &  -2.6780\\
K-207                    &13 25 56.87 &-43 00 44.4&  5.36& 21.04& 20.58& 19.63& 20.31& 0.442 &  -4.9713\\
K-209                    &13 25 57.42 &-42 53 41.6&  9.23& 19.45& 19.14& 18.34& 18.54& 0.570 &  -4.4101\\
K-216                    &13 25 58.71 &-43 04 30.7&  6.61& 19.26& 19.02& 18.49& 19.19&-0.297 &  -8.8381\\
K-224                    &13 26 11.86 &-43 02 43.2&  8.24& 21.56& 21.18& 20.02& 20.87& 0.408 &  -5.1258\\
K-235                    &13 26 20.42 &-42 59 46.4&  9.75& 21.60& 21.20& 20.45& 21.08& 0.345 &  -5.4180\\
\enddata
\tablenotetext{a}{Cluster Names:
BTC objects are candidate GCs from
\cite{hhg04II}, Minniti objects are candidate clusters from
\cite{m04}, and the remaining candidate GCs,  denoted by
the prefix K (with no radial velocity measurement) are from
\cite{pff04I}.}
\tablenotetext{b}{GCs with radial velocity measurement listed as an average weight of all previous measurements in \cite{woodley07}.}
\tablenotetext{c}{GC0233 matched both X-ray point source 55 and 56
within 1.5\arcsec, and one of them within 0.5\arcsec.} 
\tablenotetext{d}{
\cite{m04} suggest that
Minniti53 is spectroscopically confirmed from \cite{hghh92}, yet in
this study, we do not find that the position of Minniti53 matches any
previously known GC with a radial velocity measurement within a search radius of 2\arcsec.
We list this object as a candidate GC.}

\end{deluxetable}


\clearpage

\begin{thebibliography}{}

\bibitem[Angelini et al.(2001)]{angelini01} Angelini, L., Loewenstein,
M., \&
  Mushotzky, R.F. 2001, \apj, 557, L35

\bibitem[Beasley et al.(2000)]{beasley00} 
Beasley, M.~A.,  Sharples, R.~M., Bridges, T.~J., 
Hanes, D.~A., Zepf, S.~E., Ashman, K.~M., Geisler, D.\ 2000, \mnras, 318, 1249

\bibitem[Beasley et al.(2007)]{beasley07} Beasley, M. et al. 2007 \mnras,
submitted

\bibitem[Bellazzini et al(1995)]{bellazzini95} Bellazzini, M.,
  Pasquali, A. Federici, L., Ferraro, F.R., \& Fusi Pecci, F. 1995,
  \apj, 439, 687

\bibitem[Cohen et al (1998)]{cohen98} Cohen, J.~G., Blakeslee, J.~P., 
\& Ryzhov, A.\ 1998, \apj, 496, 808

\bibitem[Di Stefano et al.(2003)]{distefano03} Di Stefano, R., Kong,
A.K.H., 
VanDalfsen, M.L., Harris, W.E., Murray, S.S., \& Delain, K.M. 2003,
\apj, 599, 1067

\bibitem[Dufour et al.(1979)]{dufour79} Dufour, R.J., van den Bergh,
S.,
Harvel, C.A., Martins, D.H., Schiffer, F.H., Talbot, R.J., Talent,
D.L.,
\& Wells, D.C. 1979, \aj, 84, 284

\bibitem[Fabbiano(2006)]{fab06} Fabbiano, G. 2006, \araa, 44, 323

\bibitem[Fabbiano et al.(2007)]{fabbiano07} Fabbiano, G., Brassington,
  N.~J., Zezas, A. et al. 2007, astro-ph/0710.5126

\bibitem[Fabian et al.(1975)]{fpr75} Fabian, A.~C., Pringle, 
J.~E., \& Rees, M.~J.\ 1975, \mnras, 172, 15P 

\bibitem[Fan et al.(2005)]{fan05} Fan, Z., Ma, J.,
Zhou, X,. Chen, J., Jiang, Z., \& Wu, Z. 2005, \pasp, 117, 1236

\bibitem[Feigelson et al.(1981)]{feig81} Feigelson, E.~D., 
Schreier, E.~J., Delvaille, J.~P., Giacconi, R., Grindlay, J.~E., \& 
Lightman, A.~P.\ 1981, \apj, 251, 31 

\bibitem[Ferrarese et al.(2007)]{ferrarese07} Ferrarese, L., Mould,
  J.~R., Stetson, P.~B., Tonry, J.~L., Blakeslee, J.~P., Ajhar, E.~A.,
  2007, \apj, 654, 186

\bibitem[G\'omez et al.(2006)]{gomez06} G\'omez, M., Geisler, D., 
  Harris, W.E., Richtler, T., Harris, G.L.H., \& Woodley, K.A. 2006,
  \aap, 447, 877

\bibitem[G\'omez \& Woodley(2007)]{gw07} G\'omez, M. \& Woodley,
  K.~A. 2007, \apj, 670, L105

\bibitem[Graham \& Phillips(1980)]{gp80} Graham, J. A., \& Phillips,
M. M. 1980, \apj, 239, L97

\bibitem[Grindlay(1987)]{grindlay87} Grindlay, J.~E.\ 1987, The 
Origin and Evolution of Neutron Stars, 125, 173 

\bibitem[Grindlay(1993)]{grindlay93} Grindlay, J.~E.\ 1993, The 
Globular Cluster-Galaxy Connection, 48, 156

\bibitem[Hardcastle et al.(2007)]{hardcastle07} Hardcastle, M.~J.,
  Kraft, R.~P., Sivakoff, G.~R. et al. 2007, astro-ph/0710.1277

\bibitem[Harris et al.(1992)]{hghh92} Harris, G.L.H., Geisler, D.,
Harris, H.C., \& Hesser, J.E. 1992, \aj, 104, 613

\bibitem[Harris et al.(2006)]{harris06} Harris, W.E., Harris, G.L.H.,
Barmby, P., 
McLaughlin, D.E., \& Forbes, D.A. 2006, \aj, 132, 2187

\bibitem[Harris et al.(2006)]{har06} Harris, W.E., Whitmore, B.C.,
  Karakla, D., Oko\'n, W., Baum, W.A., Hanes, D.A., \& Kavelaars,
  J.J. 2006, \apj, 636, 90H

\bibitem[Harris, Harris \& Geisler(2004)]{hhg04II} Harris, G.L.H.,
Harris, W.E., \& Geisler, D. 2004, \aj, 128, 723

\bibitem[Harris \& Harris(2002)]{hh02} Harris, W.E. \& Harris, G.L.H.
2002, \aj, 123, 3108

\bibitem[Harris et al.(2002)]{har02} Harris, W.E., Harris, G.L.H.,
Holland, S.T., \& McLaughlin, D.E. 2002, \aj, 124, 1435

\bibitem[Haynes et al.(1983)]{haynes83} Haynes, R.~F., Cannon, 
R.~D., \& Ekers, R.~D.\ 1983, Proceedings of the Astronomical Society
of Australia, 5, 241 

\bibitem[Hills(1976)]{hills76} Hills, J.~G.\ 1976, \mnras, 175, 
1P 

\bibitem[Hui et al.(1995)]{hui95} Hui, X., Ford, H.C., Freeman, K.C.,
\& Dopita, M.A. 1995, \apj, 449, 592

\bibitem[Irwin(2005)]{irwin05} Irwin, J.~A. 2005, \apj, 631, 511


\bibitem[Israel(1998)]{israel98} Israel, F.~P.\ 1998, \aapr, 8, 
237 


\bibitem[Ivanova(2006)]{ivanova06} Ivanova, N.\ 2006, \apj, 653, L137 

\bibitem[Jord\'an (2004)]{jor04} Jord\'an, A. 2004, \apj, 613, L117

\bibitem[Jord\'an et al.(2007)]{jordan07} Jord\'an, A., Sivakoff, G.~R.,
  McLaughlin, D.~E., et al. 2007, \apj, 671, L117

\bibitem[Jord\'an et al.(2002)]{jordan02} Jord\'an, A., C\^ot\'e, P.,
West, M.~J., \& Marzke, R.~O.\ 2002, \apj, 576, L113

\bibitem[Jord\'an et al.(2004)]{jordan04} Jord\'an, A., C\^ot\'e, P.,
  Ferrarese, L., Blakeslee, J.P., Mei, S., Merritt, D.,
  Milosavljevi\'c, M., Peng, E.W., Tonry, J.L., \& West, M.J. 2004,
  \apj, 613, 279 

\bibitem[Jord\'an et al.(2005)]{jordan05} Jord\'an, A., C\^ot\'e, P.,
Blakeslee, J.P., Ferrarese, L., McLaughlin, D.E., Mei, S., Peng, E.W.,
Tonry, J.L., Merritt, D., Milosavljevi\'c, M., Sarazin, C.L., Sivakoff, G.R.,
\& West, M.J. 2005, \apj, 634, 1002

\bibitem[Juett(2005)]{juett05} Juett, A.~M. 2005, \apj, 621, L25

\bibitem[Karachentsev et al.(2007)]{kara07} Karachentsev, I.D. et al.~
2007, \aj, 133, 504

\bibitem[Katz(1975)]{katz75} Katz, J.~I.\ 1975, \nat, 253, 698 


\bibitem[Kim et al.(2006)]{kim06} Kim, E., Kim, D.-W., 
Fabbiano, G., Lee, M.~G., Park, H.~S., Geisler, D., \& Dirsch, B.\ 2006, 
\apj, 647, 276 


\bibitem[Kraft et al.(2000)]{kra00} Kraft, R.~P., et al.\ 
2000, \apjl, 531, L9 

\bibitem[Kraft et al.(2001)]{kra01} Kraft, R.P., Kregenow J.M.,
Forman, W.R., Jones, C., \& Murray, S.S.  2001, \apj, 560, 675

\bibitem[Kraft et al.(2002)]{kra02} Kraft, R.~P., Forman, 
W.~R., Jones, C., Murray, S.~S., Hardcastle, M.~J., \& Worrall, D.~M.\ 
2002, \apj, 569, 54 

\bibitem[Kraft et al.(2003a)]{kra03a} Kraft, R.~P., Hardcastle, 
M.~J., Forman, W.~R., Jones, C., Murray, S.~S., \& Worrall, D.~M.\ 2003, 
New Astronomy Review, 47, 625

\bibitem[Kraft et al.(2003b)]{kra03b} Kraft, R.~P., 
V{\'a}zquez, S.~E., Forman, W.~R., Jones, C., Murray, S.~S.,
Hardcastle, 
M.~J., Worrall, D.~M., \& Churazov, E.\ 2003, \apj, 592, 129 


\bibitem[Kundu \& Whitmore(2001)]{kundu01} Kundu, A., \& Whitmore,
 B.C. 2001, \aj, 121, 2950

\bibitem[Kundu et al.(2002)]{kundu02} Kundu, A., Maccarone, T.J., \&
  Zepf, S.E. 2002, \apj, 574, L5

\bibitem[Kundu et al.(2003)]{kundu03} Kundu, A., Maccarone, T.J.,
  Zepf, S.E., \& Puzia, T.H. 2003, \apj, 589, L1 

\bibitem[Kundu et al.(2007)]{kundu07} Kundu, A., Maccarone, T.~J.,
\& Zepf, S.E.  2007, \apj, 662, 525

\bibitem[Maccarone et al.(2004)]{maccarone04} Maccarone, T.~J., 
Kundu, A., \& Zepf, S.~E.\ 2004, \apj, 606, 430 

\bibitem[McLaughlin(2000)]{mclaughlin00} McLaughlin, D.E. 2000, \apj,
539, 618

\bibitem[McLaughlin et al.(2007)]{mcl07} McLaughlin, D.E., Barmby, P.,
Harris, W.E., Harris, G.L.H., \& Forbes, D.A. 2007, \mnras, submitted

\bibitem[Minniti et al.(2004)]{m04} Minniti, D., Rejkuba, M., Funes,
  J.~G., \& Akiyama, S. 2004, \apj, 600, 716

\bibitem[Moretti et al.(2003)]{moretti03} Moretti, A., Campana, 
S., Lazzati, D., \& Tagliaferri, G.\ 2003, \apj, 588, 696

\bibitem[Peng, Ford, \& Freeman(2004a)]{pff04I} Peng, E.W., Ford, H.C.,
\& Freeman, K.C. 2004a, \apjs, 150, 367

\bibitem[Peng, Ford, \& Freeman(2004b)]{pff04II} Peng, E.W., Ford, H.C.,
\& Freeman, K.C. 2004b, \apj, 602, 705

\bibitem[Peng et al.(2002)]{peng02} Peng, E.W., Ford, H.C., Freeman,
K.C., \& White, R.L. 2002, \aj, 124, 3144

\bibitem[Peng et al.(2006)]{peng06} Peng, E.W., Jord\'an, A., C\^ot\/e,
P., Blakeslee, J.P., 
Ferrarese, L., Mei, S., West, M.J., Merritt, D., Milosavljevi\/c, M.,
\& Tonry, J.L.
2006, \apj, 639, 95

\bibitem[Pooley et al.(2003)]{pooley03} Pooley, D. et~al.\ 2003,
\apj, 591, L131

\bibitem[Posson-Brown et al.(2007)]{poss07} Posson-Brown, J., 
Raychaudhury, S., Forman, W., Hank Donnelly, R., \& Jones, C.\ 2007, 
submitted to \apj (astro-ph/0605308)

\bibitem[Puzia et al (1999)]{puzia99} Puzia, T~H.,
Kissler-Pattig, M., Brodie, J.~P., \& Huchra, J.~P. 1999, \aj, 118, 2734

\bibitem[Raychaudhury(1989)]{ray89} Raychaudhury, S.\ 1989, 
\nat, 342, 251 
 
\bibitem[Rejkuba(2001)]{rej01} Rejkuba, M. 2001, \aap, 369, 812

\bibitem[Rejkuba et al.(2005)]{rej05} Rejkuba, M., Greggio, L.,
Harris, W.E., Harris, G.L.H., \& Peng, E.W. 2005, \apj, 631,262


\bibitem[Rosenberg et al.(1999)]{rosenberg99} Rosenberg, A., Saviane, I., 
Piotto, G., \& Aparicio, A. 1999, \aj, 118, 2306

\bibitem[Sarazin et al.(2001)]{sarazin01} Sarazin, C.~L., Irwin, 
J.~A., \& Bregman, J.~N.\ 2001, \apj, 556, 533 


\bibitem[Sarazin et al.(2003)]{sarazin03} Sarazin, C.L., Kundu, A.,
  Irwin, J.A., Sivakoff, G.R., Blanton, E.L., \& Randall, W.E. 2003,
  \apj, 595, 743

\bibitem[Schiminovich et al.(1994)]{schim94} Schiminovich, D., 
van Gorkom, J.~H., van der Hulst, J.~M., \& Kasow, S.\ 1994, \apjl,
423, 
L101 

\bibitem[Sivakoff et al.(2007)]{siv07} Sivakoff, G.R., Jord\'an, A.,
Sarazin, C.L.,
Blakeslee, J.P., C\^ot\'e, P., Ferrarese, L., Juett, A.M., Mei, S., \&
 Peng, E.W.
2007, \apj, 660, 1246

\bibitem[Turner et al.(1997)]{turner97} Turner, T.~J., George, 
I.~M., Mushotzky, R.~F., \& Nandra, K.\ 1997, \apj, 475, 118 

\bibitem[Voss \& Gilfanov(2006)]{vg06} Voss, R., \& 
Gilfanov, M.\ 2006, \aap, 447, 71 


\bibitem[Voss \& Gilfanov(2007)]{voss07} Voss, R., \& 
Gilfanov, M.\ 2007, \aap, 468, 49 

\bibitem[White \& Angelini(2001)]{white01} White, N. E., \& Angelini,
L. 2001, 
\apj, 561, L101 

\bibitem[White et al.(2002)]{white02} White, R.~E., III, 
Sarazin, C.~L., \& Kulkarni, S.~R.\ 2002, \apjl, 571, L23 

\bibitem[Woodley et al.(2005)]{WHH05} Woodley, K.A., Harris, W.E., 
\& Harris, G.L.H. 2005, \aj, 129, 2654

\bibitem[Woodley(2006)]{woodley06} Woodley, K.A. 2006, \aj, 132, 2424

\bibitem[Woodley et al.(2007)]{woodley07} Woodley, K.~A., Harris, 
W.~E., Beasley, M.~A., Peng, E.~W., Bridges, T.~J., Forbes, D.~A., \& 
Harris, G.~L.~H.\ 2007, \aj, 134, 494

\bibitem[Worthey(1994)]{worthey94} Worthey, G.\ 2004, \apjs, 95, 107

\end{thebibliography}
\end{document}